\documentclass{article}
\usepackage{graphicx} 
\usepackage[square, numbers, comma, sort&compress]{natbib}
\usepackage{booktabs}
\usepackage{caption}
\usepackage{subcaption}
\usepackage{multirow}
\usepackage{makecell}
\usepackage{amsmath}
\usepackage{siunitx}
\usepackage{dcolumn}
\usepackage[table]{xcolor}
\usepackage{pifont}
\usepackage{amssymb}
\usepackage{ragged2e}
\usepackage{url} 
\usepackage[a4paper, margin=1in]{geometry}
\usepackage[acronym]{glossaries}
\usepackage{abbr}

\title{Minimally Invasive Brain Computer Interfaces: Evaluating the Impact of Tissue Layers on Signal Quality of Sub-Scalp EEG}
\author{T B Mahoney, J Liu, H Xin, D B Grayden, S E John}
\date{December 2024}

\begin{document}

\maketitle

\section*{Abstract}

Individuals with severe physical disabilities often experience diminished quality of life stemming from limited ability to engage with their surroundings. Brain-Computer Interface (BCI) technology aims to bridge this gap by enabling direct technology interaction. However, current BCI systems require invasive procedures, such as craniotomy or implantation of electrodes through blood vessels, posing significant risks to patients. Sub-scalp electroencephalography (EEG) offers a lower risk alternative. This study investigates the signal quality of sub-scalp EEG recordings from various depths in a sheep model, and compares results with other methods: ECoG and endovascular arrays. A computational model was also constructed to investigate the factors underlying variations in electrode performance. We demonstrate that peg electrodes placed within the sub-scalp space can achieve visual evoked potential signal-to-noise ratios (SNRs) approaching that of ECoG. Endovascular arrays exhibited SNR comparable to electrodes positioned on the periosteum. Furthermore, sub-scalp recordings captured high gamma neural activity, with maximum bandwidth ranging from 120~Hz to 180~Hz depending on electrode depth. These findings support the use of sub-scalp EEG for BCI applications, and provide valuable insights for future sub-scalp electrode design. This data lays the groundwork for human trials, ultimately paving the way for chronic, in-home BCIs that empower individuals with physical disabilities.

\section{Introduction}
Neurological disorders are currently the leading cause of disability globally, and have increasing incidence as a result of population growth and ageing \citep{feigin_global_2020}. Particularly severe conditions such as motor neurone disease, with a global incidence of 64,000 \citep{logroscino_global_2018}, can leave patients unable to move or communicate, resulting in a significant reduction in quality of life. Brain-computer interfaces (BCIs) aim to establish a direct communication link between neural activity in the brain and an external device. For people with severe neurological conditions, BCIs offer improved quality of life by facilitating communication and interaction with the world around them \citep{chaudhary_chapter_2016,khan_review_2020, kubler_user-centered_2014, mcfarland_brain-computer_2020}. 

Current BCI technology has several limitations. Invasive implants, such as electrocorticography (ECoG) and penetrating arrays, are accompanied by significant risks to the patient during both implantation and chronic use, and can result in scaring that damages neural tissue and attenuates signal quality over time \citep{johnston_complications_2006, nagahama_intracranial_2019, onal_complications_2003, rolston_national_2015, taussig_invasive_2012, winslow_comparison_2010, winslow_quantitative_2010, theunisse_risk_2018}. Non-invasive electroencephalography (EEG) involves lengthy donning and doffing procedures and inconsistent signal stability. Most people regard EEG as unaesthetic and cumbersome by potential BCI users \citep{kubler_user-centered_2014, miralles_braincomputer_2015, rashid_current_2020}. Endovascular (EV) stent-electrode arrays have recently attempted to address these limitations \citep{oxley_motor_2021}; however, these devices have poor spatial coverage (being confined to brain regions with large vascular structures), cannot be removed once implanted, and may lead to adverse effects seen with similar stenting procedures \citep{starke_endovascular_2015, modi_stent_2024}. There is a need for a minimally invasive signal acquisition method that addresses the disadvantages with current technologies and that will enable widespread long-term BCI use.

Placement of electrodes into the sub-scalp (also termed sub-galeal or sub-dermal) space offers a viable solution for BCI that overcomes many of the disadvantages of current signal acquisition methods.  Unlike EEG caps, sub-scalp devices are implanted under the skin providing improved stability, are discrete, and require no donning and doffing procedures. They do not require brain or vascular surgery to be implanted. Sub-scalp electrodes can be positioned as needed across the skull, providing access to activity from brain regions outside the reach of EV arrays and can be easily removed if needed for replacement or upgrades.  The safety of chronic implantation of electronics beneath the scalp has been demonstrated by cochlear implants around the world over decades \citep{jiang_analysis_2017, weder_management_2020, desasouza_cochlear_2022}. Sub-scalp EEG is currently being employed for seizure monitoring and has shown promise for long-term use \citep{duun-henriksen_new_2020, stirling_seizure_2021, weisdorf_ultra-long-term_2019}. Current evidence of safety and stability provides hope for sub-scalp BCIs to record high-quality neural signal for decoding and control of external devices. 

sub-scalpElectrodes can be placed either above the periosteum (a layer of tissue covering the skull), directly on the skull, or partially embedded within the skull (referred to as peg electrodes). There is a trade-off between electrode depth and signal quality. As the electrode is placed closer to the source location, there is less signal attenuation; however, the procedure to place the electrode may be longer, more complicated, and more traumatic, which increases risk and cost to the user, as well as recovery time. For example, ring electrodes (electrodes lining a silicone shaft) can be tunnelled to distal locations across the skull from a small incision in a simple day procedure, but they may record with reduced signal quality due to the periosteum and skull layers. On the other hand, peg electrodes require a bur hole in the skull at each site, but may record with much higher quality signal. The optimal placement of the device within the sub-scalp space is not known. 

Sub-scalp EEG devices aim to find a minimally invasive option that allows for signal quality that is sufficiently high for the intended application. A study by \citet{benovitski_ring_2017} investigated differences between ring, peg, and disc electrodes in the sub-scalp space concerning amplitude-integrated EEG (aEEG), chewing artefact, impedance, and histology after 3-6 months of implantation in sheep. The study concluded that peg electrodes compared with disc and ring electrodes saw significantly attenuated chewing artefact (by a factor of 3 and 6, respectively) and higher aEEG (by a factor of 4 and 5, respectively). While these results provide valuable insights, the quality of neural activity used to control BCI devices, such as visual evoked potentials (VEPs), as recorded from sub-scalp electrodes is unknown. 

Prior research comparing signal quality of sub-scalp EEG with other neural recording methods is limited. Sub-scalp EEG has demonstrated capture of high gamma activity, though with less power than ECoG \citep{olson_comparison_2016}. Preliminary work has indicated that sub-scalp EEG may have similar maximum bandwidth to endovascular arrays \citep{mahoney_comparison_2023}. Regarding signal-to-noise ratio (SNR), previous work has demonstrated significantly higher SNR in ECoG recordings than surface EEG \citep{ball_signal_2009, wittevrongel_decoding_2018}, and similar SNR between epidural ECoG and endovascular stent arrays \citep{john_signal_2018}. Regarding the different layers within the sub-scalp space, we expect EEG SNR and maximum bandwidth will increase as the sub-scalp electrodes are placed closer to the surface of the brain. However, the degree of change and the impact of each layer individually are not known. 

The aim of this study is to quantify and compare the SNR and maximum bandwidth of sub-scalp EEG recorded above the periosteum, on the skull surface, and partially within the skull (peg). Additionally, we compare these sub-scalp recordings with current gold standard BCI signal acquisition methods: electrocorticography (ECoG) and endovascular electrodes, to characterise the sub-scalp methods within the neural signal acquisition space. We do this both with visual evoked potentials stimulated in sheep models, and through simulations. Results from these experiments will provide support for long-term, in-human sub-scalp BCI clinical studies that hopefully propel this technology toward regulatory approval and availability for persons with severe paralysis, providing them with a means of interacting with their environment. 

\section{Methods}

\subsection{Experiment Procedure}
\begin{figure}
    \centering
    \sbox0{\begin{subfigure}[b]{0.6\linewidth}
        \includegraphics[width=\linewidth]{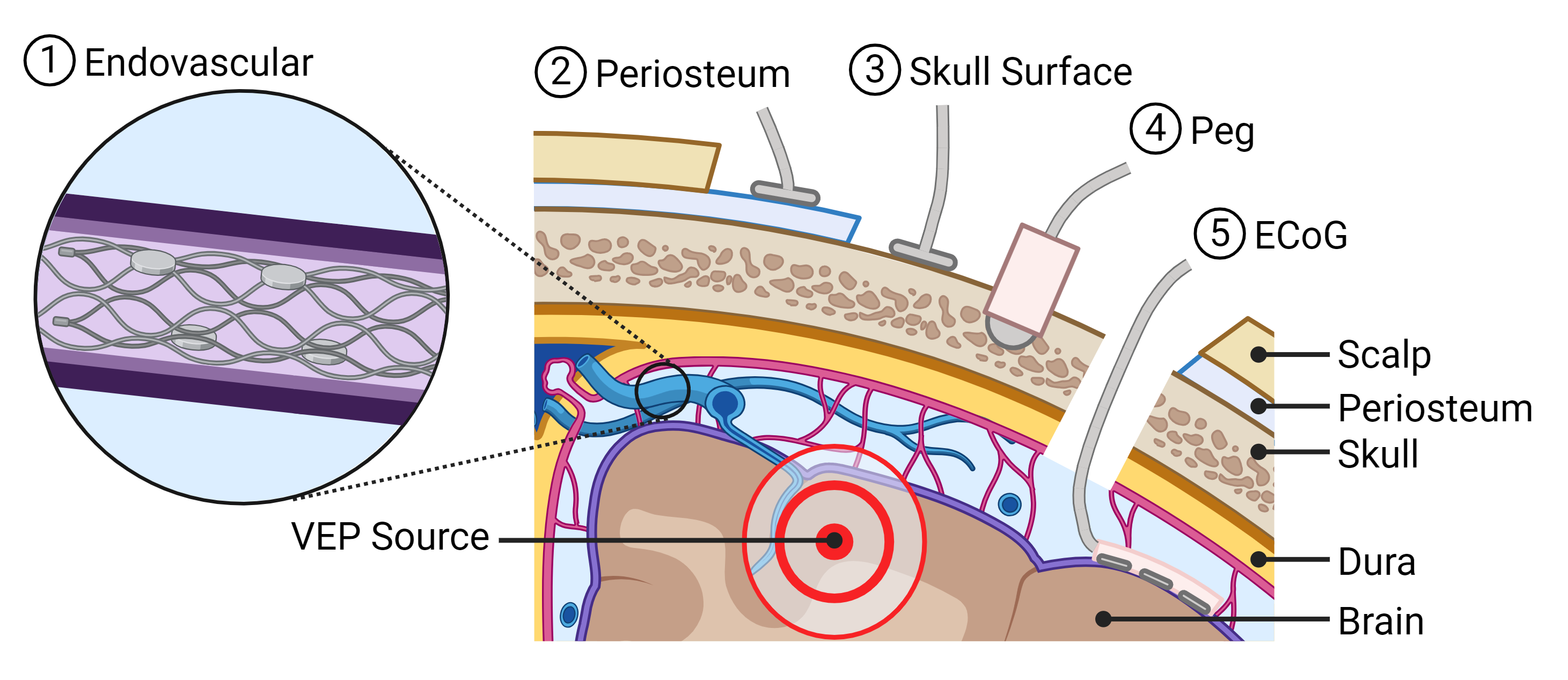}
        \caption{}
        \label{fig:CH3_Fig1_a}
    \end{subfigure}}%
    \usebox0\hfill\begin{minipage}[b][\ht0][s]{0.39\linewidth}
        \begin{subfigure}{0.49\linewidth}
            \includegraphics[width=\linewidth]{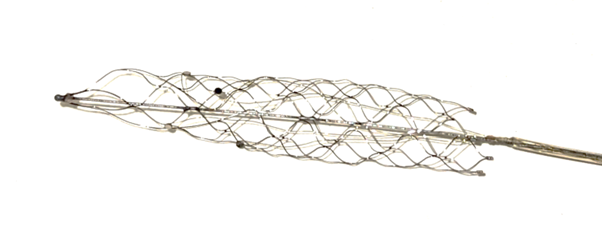}
            \caption{}
            \label{fig:CH3_Fig1_b}
        \end{subfigure}
        \hfill
        \begin{subfigure}[b]{0.49\linewidth}
            \includegraphics[width=\linewidth]{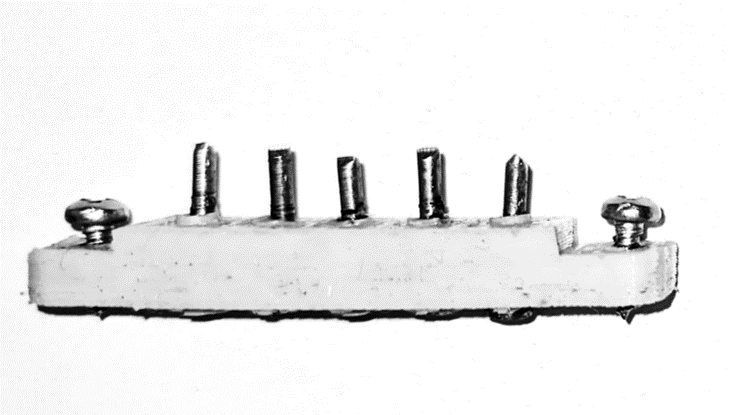}
            \caption{}
            \label{fig:CH3_Fig1_c}
        \end{subfigure}
        \vfill
        \begin{subfigure}{0.49\linewidth}
            \includegraphics[width=\linewidth]{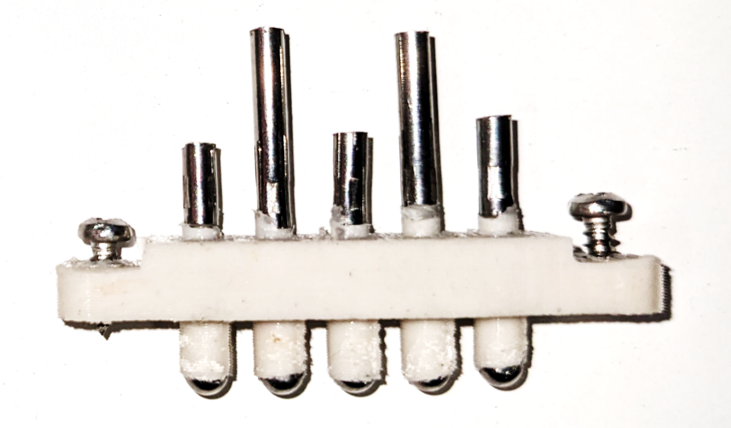}
            \caption{}
            \label{fig:CH3_Fig1_d}
        \end{subfigure}
        \hfill
        \begin{subfigure}[b]{0.49\linewidth}
            \includegraphics[width=\linewidth]{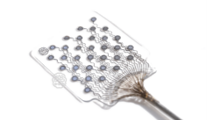}
            \caption{}
            \label{fig:CH3_Fig1_e}
        \end{subfigure}
    \end{minipage}
    \vspace{0.5cm}
    
    \begin{subfigure}[b]{\linewidth}
        \centering
        \includegraphics[scale=0.4]{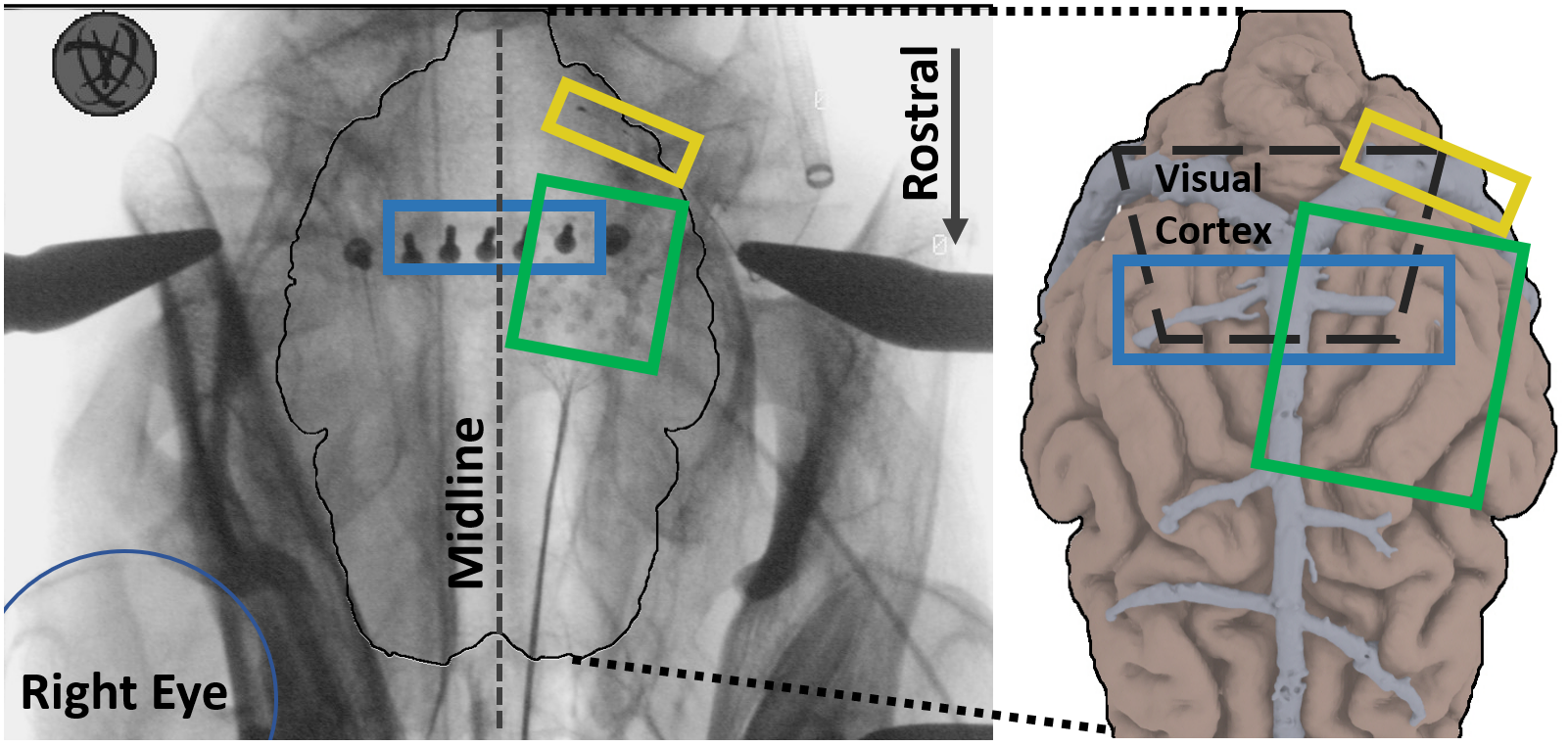}
        \caption{}
        \label{fig:CH3_Fig1_f}
    \end{subfigure}
    \caption[Electrodes and Locations Diagram]{Electrodes and their placements. (a) Electrode depth locations, showing five different depth placements. Unlike the illustration, electrodes were vertically aligned as closely as possible (created with BioRender.com). Images of each electrode type used: (b) the endovascular array, (c) periosteum and skull surface, (d) peg, and (e) ECoG. (f) X-ray image showing electrode locations in the sheep near the visual cortex. The blue rectangle outlines the location of the periosteum, skull surface and peg electrodes (skull surface in this case), the green rectangle outlines the subdural ECoG array, and the yellow rectangle outlines the endovascular array (contralateral to the side of stimulus). The array positions are also superimposed over a sheep brain model to the right of the X-ray image (image source: \citet{oxley_minimally_2016}); the visual cortex is outlined by the dashed trapezoid.}    
\end{figure}

This experiment was approved by the Animal Ethics Committee at the Florey Institute of Neuroscience and Mental Health, Melbourne, Australia (Approval number: 22010). Six female Corriedale sheep were used for this study. The sheep were anaesthetised with isoflurane. Electrodes were sequentially placed at five different recording sites, as illustrated in Figure \ref{fig:CH3_Fig1_a}. The locations were:
\begin{enumerate}
    \item \textbf{Endovascular:} An endovascular stent-electrode array (Figure \ref{fig:CH3_Fig1_b}) was deployed in the transverse sinus through the jugular vein, contralateral to the side of stimulus. The array included four electrodes (ø750~µm).
    
	\item \textbf{Periosteum:} Upon completion of the endovascular recording session, the scalp was removed, revealing the periosteum over the skull. A five-channel array of disc-like stainless steel electrodes set in silicone (ø3~mm, 5~mm pitch, Figure \ref{fig:CH3_Fig1_c}) were sutured into the periosteum over the occipital bone, approximately 5~mm caudal to the lambdoid suture. 
 
	\item \textbf{Skull Surface:} Upon completion of periosterum recording, the periosteum was removed to reveal the skull. The same electrode array was then screwed into the skull at the same location as during the periosteum recording. 
 
	\item \textbf{Peg:} Upon completion of the skull surface recording, a bur hole was made directly beneath the electrode array, spanning the length of all five channels, with a depth of 4~mm. An array with was inserted into the cavity and screwed into place (ø3~mm, 5~mm pitch, Figure \ref{fig:CH3_Fig1_d}). 
 
	\item \textbf{ECoG:} Upon completion of the peg recordings, a section of skull was removed and an ECoG array (AirRay, Cortec, Germany, Figure \ref{fig:CH3_Fig1_e}) was inserted into the subdural space, below the previous recording site, over the visual cortex. Six ECoG channels were chosen based on their proximity to both the sub-scalp electrode recording site and the endovascular array position. 
\end{enumerate}

Figure \ref{fig:CH3_Fig1_f} shows an x-ray of the arrays \textit{in vivo} and their estimated position over the visual cortex. A reference electrode was placed in the rostral sub-scalp space, distal from the occipital bone. The animal was euthanised at the conclusion of the experiment via lethabarb injection.

\subsection{Visual Stimulus Procedure}
The following procedures were performed for each electrode location. Room lights were turned off and blinds were closed to minimise ambient light. A full field flash stimulator (Grass Instrument Co., USA) was placed approximately 20~cm from the right eye. The animal’s right eye was taped open and saline was applied to the eye at regular intervals to maintain eye health. The stimulator provided a single flash stimulus at 0.99~Hz for 5 minutes. A photodiode connected to an analog-to-digital converter (Arduino Nano 33 BLE, Arduino, Italy) detected the flash and notified the BLE receiver python script on the receiver laptop (EliteBook x360 1030 G8 Notebook, HP, USA), acting as a trigger channel. A 4~min background recording was taken in the dark without stimulus and used for power spectrum analysis. Recordings were sampled with an electrophysiology amplifier chip (RHD2132, Intan Technologies, U.S.A.), with a sampling frequency of 1024~Hz. 

\subsection{Analysis Methods}
Signal quality was quantified in terms of VEP SNR and maximum bandwidth. All analyses were conducted in MATLAB (Version 2023a, MathWorks, USA).

\subsubsection{VEP Amplitude and SNR}
A key metric for signal quality of neural recording methods is SNR. A method of computing SNR in EEG, for which it is often difficult to differentiate signal from noise, is by evoking a response to a stimulus. Visual evoked potentials (VEPs) are a response seen in the brain due to a visual stimulus. VEP detection is particularly useful for BCI applications that rely on the user attending to a visual stimulus mapped to a command.

The raw data were bandpass filtered between 5-40~Hz using MATLAB's bandpass function. This function uses a minimum-order with stopband attenuation of 60 dB and includes delay compensation. Data were then segmented into epochs about the time of stimulus. Epochs with a range greater than 1~mV were considered contaminated with artefact and were rejected. The amplitude of the VEP was calculated by averaging the response over epochs and calculating the peak-to-peak voltage between the time of stimulus and 800~ms post stimulus. Signal-to-noise (SNR) ratio was calculated as the ratio of the variance of the signal 300~ms post-stimulus to the variance of the signal 300~ms pre-stimulus. These windows were used because the VEP waveform was observed to span 300~ms post-stimulus. As saline was applied to the eye between sessions, a random sample of 50 trials were used for analysis to account for any variation in response over the 5~min of recording. The two channels that exhibited the highest SNR were included in the analysis for each recording method.

\subsubsection{Bandwidth}
Neural activity often presents as oscillations. As such, the frequency domain contains useful features for classification of brain states. Neural power exponentially decreases toward the noise floor as frequency increases \citep{miller_power-law_2009}. Due to this decay, EEG components in the high gamma band ($\geq$70~Hz), which are particularly useful features for classifying brain states, can be difficult to extract when recording from outside the body, where brain signals have been significantly attenuated.

\begin{figure}[ht]
\includegraphics[scale=0.4]{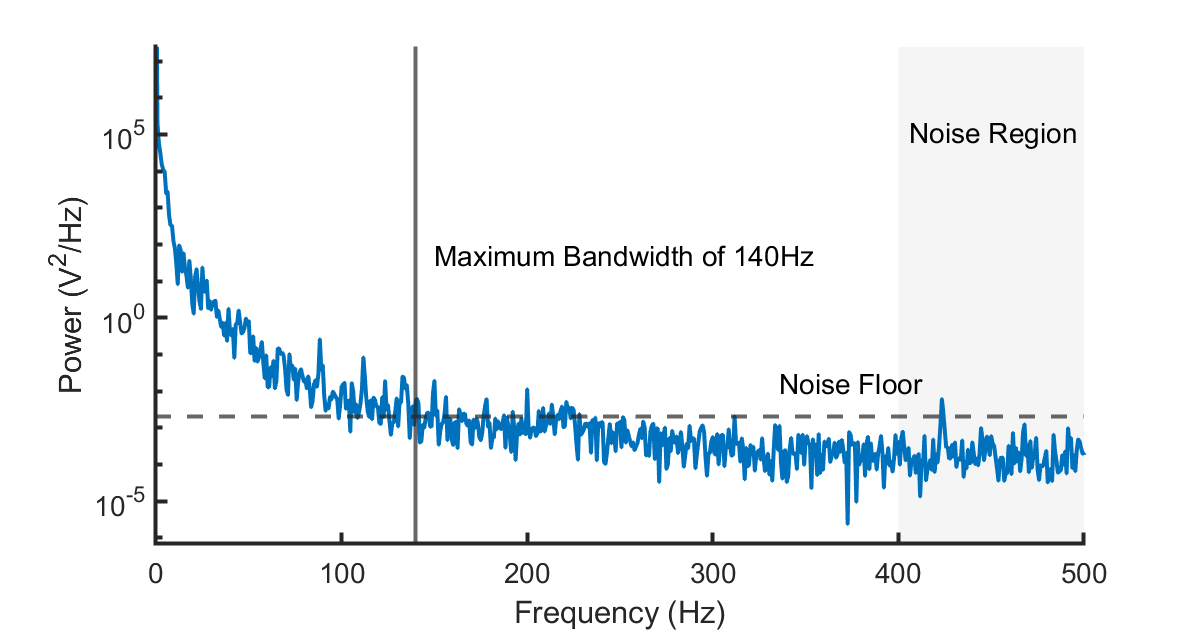}
\centering
\caption[Maximum Bandwidth Calculation Diagram]{A typical power spectrum of EEG data. The frequency at which the power drops below the noise floor (dashed line) is considered the maximum bandwidth, indicated by the solid vertical line. This example was recorded using a peg electrode in sheep 3.}
\label{fig:CH3_Fig2}
\end{figure}

We can estimate the maximum measurable frequency of neural activity that each recording location is capable of capturing. Background neural recordings of 4~min duration were used and only the two channels resulting in the highest VEP SNR were considered. The recordings were lowpass filtered with a 500~Hz cutoff frequency. The power spectrum was computed using MATLAB's pwelch function. Power in the 400-500~Hz band was considered the noise floor, though frequencies ±10~Hz about 50~Hz harmonics were not included. A conservative noise floor threshold was calculated by adding the third quartile of the noise floor band power with 1.5 times the interquartile range (IQR), an approach that has been used previously a similar study \citep{john_signal_2018}. Bins spanning 10~Hz were computed based on median power. Scanning through the bins from 10 to 400~Hz, the bin power prior to the first bin to drop below the computed noise floor was considered the maximum bandwidth (Figure~\ref{fig:CH3_Fig2}). 

\subsection{Forward Simulation of EEG signals}
\begin{figure}[htb]
    \vspace{-\baselineskip}
    \centering
    \includegraphics[width=0.6\linewidth]{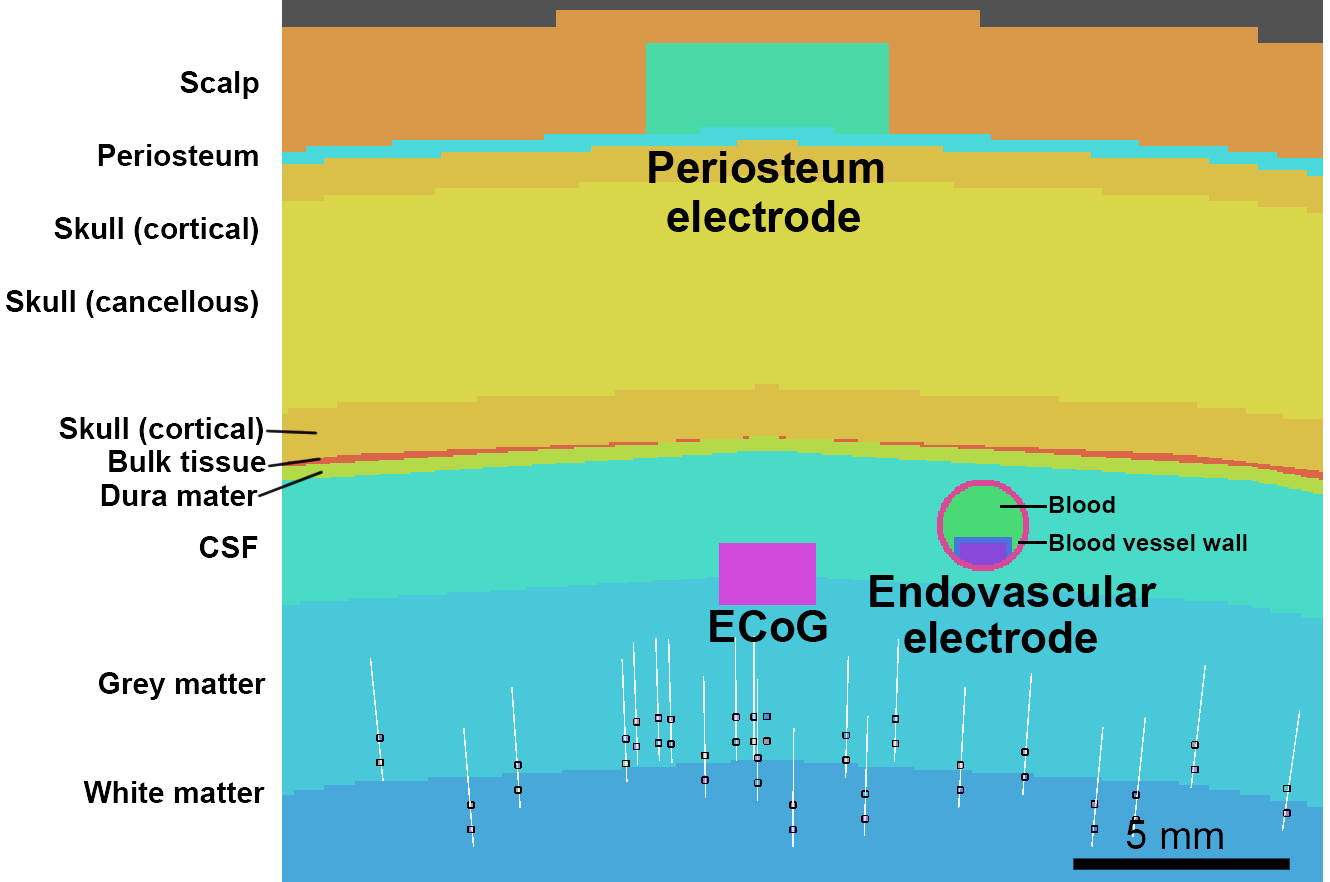}
    \caption{Cross-sectional view of the voxelated \Gls{fem} model in the region of interest, illustrating the spatial configuration of the periosteum electrode, endovascular electrode at 0°, and ECoG electrode. Tissue layers and electrode placements are represented with their respective boundaries. Dipole-equivalent potential source cubes and their orientations are shown within the grey matter}
    \vspace{1em}
    \label{cross-sec}
\end{figure}

The recording performance of various electrode depths was evaluated by solving the EEG forward problem using the Finite Element Method (\gls{fem}). Simulations of the electromagentic field within a simplified volume conduction model of the ovine head were performed with Sim4Life V8.2.0 (ZMT, Zurich, Switzerland). The simplified ovine brain was modelled as a sphere deformed to match the radii in the three axes of a population-averaged brain atlas \citep{nitzsche_stereotaxic_2015}. The whole head was modelled as a sphere with the equivalent sphere radius of corriedale sheep (8.62~cm) \citep{ormachea_principal_2022}. The tissue conductivities were assumed to be isotropic and homogeneous within each tissue layer. Within the frequency ranges of interest of sub-scalp and intracanial EEG, the quasi-static approximation of the electromagnetic field holds. Tissue conductivities were assumed to be independent of frequency and the capacitive effects were neglected. Electrical properties from the IT'IS low-frequency tissue database \citep{itis_foundation_tissue_2024} were assigned to each tissue volume, including scalp with a conductivity 0.439~S/m, the inner and outer cortical bone layers of the skull (0.006~S/m), cancellous layer of the skull (0.100~S/m), the dura mater (0.060~S/m), the cerebrospinal fluid (1.879~S/m), the grey matter (0.419~S/m), and the white matter (0.348~S/m). In addition, a blood vessel with a lumen diameter of 1.3~mm and a wall thickness of 0.1~mm was modelled in the subdural space to accomodate the endovascular electrode array. The blood and blood vessel wall were assigned conductivities of 0.662~S/m and 0.232~S/m, respectively. Due to the variability in periosteum thickness and the lack of its tissue-specific electrical properties in literature, thin layer of periosteum resembling the thickness (0.23~mm) and electrical properties of the dura mater was added on the exterior surface of the skull \citep{kinaci_histologic_2020}. Finally, the bulk tissue of the head was assigned with a generic bulk conductivity of 0.33~S/m \citep{neufeld_functionalized_2016}. Figure \ref{cross-sec} demonstrates a cross-sectional view of the tissue conductor model.

The various electrodes were modelled with a close resemblance to those used in the in vivo experiments including a subdural disk electrode (ECoG electrode, 1.5~mm diameter) placed directly on the grey matter, a spherical peg electrode (Skull surface electrode, 3~mm diameter) placed partially embedded in the skull to a depth of 4 mm, and a disc electrode (3 mm diameter) underneath (skull surface electrode) and on top of the periosteum (periosteum electrode). In addition, endovascular disc electrodes (Endovascular 0°, 90°, and 180°, 0.75 mm diameter) were placed within the blood vessel lumen conforming to the blood vessel wall, oriented at 0, 90, and 180 degrees relative to the grey matter. The subdural, peg, skull surface,and periosteum electrodes were cocentric along the z-axis, while the blood vessel was offset by 3.5 mm from the centre of the head to avoid overlap with the subdural electrode. All electrodes were insulated with a 0.1 mm layer of silicone backing with a conductivity of $10^{-12}$~S/m so that only one surface (half the spherical surface in the case of the peg electrode) was exposed to tissue. A large remote return electrode was modelled as a cylinder with a diameter of 6 mm was placed in the frontal region of the head. All electrodes were assigned perfect electric conductor properties, which assume uniform potential within the conductor and zero tangential electric fields at its surface.

Cortical current dipoles were modelled as small cubic volumes with fixed potential boundary conditions. Fifty dipoles were randomly sampled within the grey matter, oriented normal to the boundary between the grey matter and white matter at a 0.5~mm distance from it. These dipoles were confined within a circular region with a diameter of 22.5~mm, approximating the dimensions of the sheep visual cortex \citep{clarke_cortical_1976}. Each dipole was approximated by a pair of cubes (edge length 0.1~mm) 0.4~mm apart centre-to-centre, which were aligned to the z-axis for convenience during the voxelation step. The cubes were assigned with equal potentials opposite in polarity approximated with the equation,
\begin{equation}
V = \frac{\left| \vec{M} \right| \cos(\theta)}{4\pi \sigma r^{2}}
\end{equation}
where $\vec{M}$ is the dipole moment, $r$ is the distance from the dipole to the point of observation, $\theta$ denotes the angle between this vector and the dipole vector, and $\sigma$ is the conductivity of the homogeneous medium \citep{schimpf_dipole_2002}. An arbitrary dipole moment of $10^{-6}$~A$\cdot$m was used for all dipoles as only the relative amplitudes of EEG recordings were of interest. Each dipole cube was further discretised with a grid size of 0.01~mm, and the average equivalent potential calculated at the grid nodes was assigned to the cube. One dipole aligned with the centre of the subdural, peg, and periosteum electrode was always imposed.

The head model was discretised with a global grid size of up to 2~mm, which was refined to up to 0.1 mm within the bounding box of all dipole locations and electrodes,  and up to 0.025~mm near the endovascular electrodes to minimise the staircase error. This led to a total of 85 million reticular voxels. A zero-flux boundary condition was also applied to the outer boundary of the simulation domain. The Electro Ohmic Quasi-Static solver of Sim4Life was used to solve for the potential field at all voxel nodes. Finally, the static recorded potential as a result of each current dipole was calculated as the potential difference between the passive potential of the recording electrode and the return electrode.

To investigate the effects of dipole location on EEG amplitude, five random distributions of fifty dipoles were generated. As the quasi-static simulation yields a static potential field, the temporal component of EEG was re-introduced by scaling with the static potential contributed by each dipole a generic exponential equation that reflects the decay of an \gls{epsp} over time,
\begin{equation}
V(t) = V_0 \cdot e^{-\frac{t}{\tau}}
\end{equation}
where $V_0$ is the resting postsynaptic potential set to 0~V for simplicity, and $\tau$ is the time constant set to 10~ms for a slow decay. The latency of \gls{epsp} of each dipole was set to be linear with the x-coordinate of the dipole presuming a propagation speed of 0.3~m/s to temporally disperse the contribution of individual dipoles with spatial dependency \citep{scheuer_velocity_2023}. The simulated recordings on different electrodes were a linear superposition of the temporally adjusted \gls{epsp} of each dipole scaled by the corresponding static recorded potential. A total of 100~ms of simulated recordings were generated as all \gls{epsp}s decayed to baseline level by the end of this time window.

\section{Results}
\label{sec:CH2_Results}

\subsection{Visual Stimulus}
\begin{figure}
    \centering
    \begin{subfigure}[b]{0.61\textwidth}
        \includegraphics[width=\textwidth]{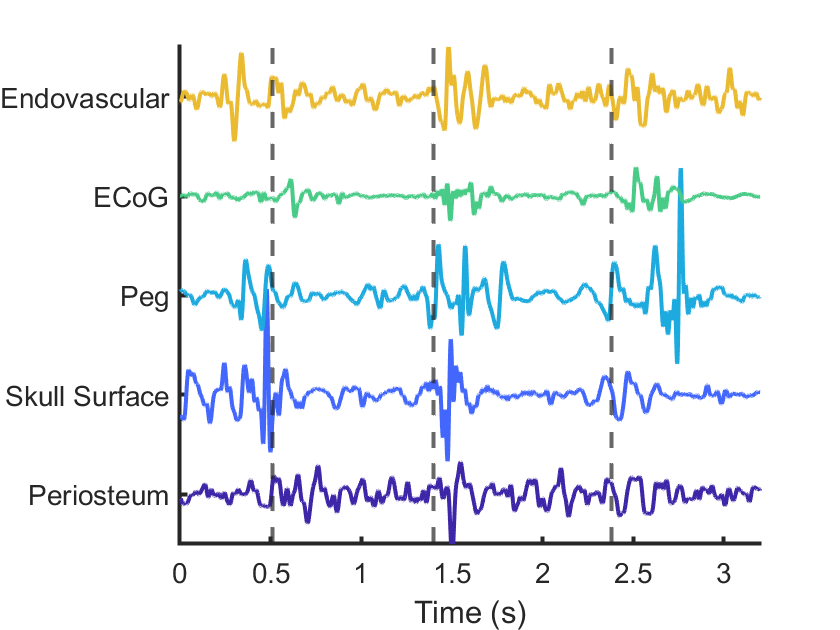}
        \caption{}
        \label{fig:CH3_Fig3_a}
    \end{subfigure}  
    \hfill
    \begin{subfigure}[b]{0.38\textwidth}
        \includegraphics[width=\textwidth]{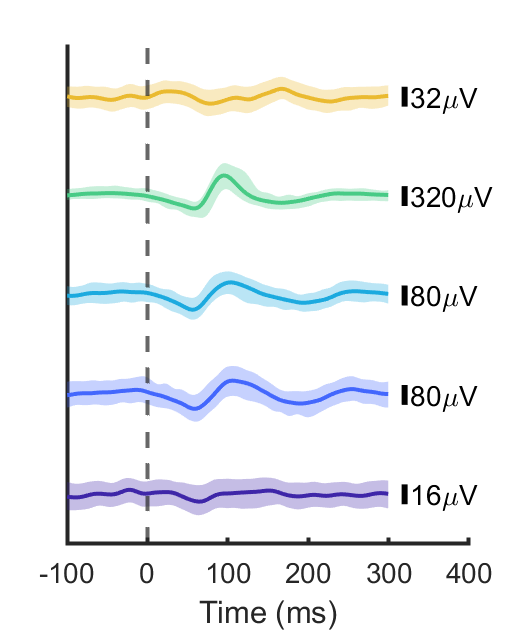}
        \caption{}
        \label{fig:CH3_Fig3_b}
    \end{subfigure} 
    \caption[Raw VEP Traces and Averaged-Over-Trials VEPs]{Examples of recordings of visual evoked responses. (a) Single traces bandpass filtered between 5-40~Hz. The dashed lines indicate the times of flashes. (b) Averaged VEP over trials. Amplitude scales for both figures are shown on the right of (b). The shaded area indicates a confidence interval of one standard deviation. Both (a) and (b) are examples taken from sheep 2.}
\end{figure}

Skull surface, peg and ECoG electrodes were recorded from in all six sheep, periosteum recordings were performed in five sheep and endovascular arrays were successfully deployed in four sheep, as summarised in Table \ref{tab:CH3_Tab1}. VEP responses were visible with all electrode types. Figure \ref{fig:CH3_Fig3_a} shows typical traces for each method from one animal. The VEP is visible in the raw traces of each electrode location. Averaging over trials reveals the underlying VEP more clearly, as shown in Figure \ref{fig:CH3_Fig3_b}. 

\begin{table}[ht]
\centering
\caption[Recording Locations Across Sheep]{Recording Locations Across Sheep. The symbols are used to indicate sheep in subsequent figures.}
\label{tab:CH3_Tab1}
\begin{tabular}{@{}lcccccc@{}}
\toprule
                   & \multicolumn{6}{c}{Sheep} \\ \cmidrule(l){2-7} 
Electrode Location & 1  & 2  & 3  & 4  & 5 & 6 \\ \midrule
Periosteum         &  & + & $*$  & $\times$  & $\square$ & $\Diamond$ \\
Skull Surface      & $\bigcirc$ & + & $*$  & $\times$  & $\square$ & $\Diamond$ \\
Peg                & $\bigcirc$ & + & $*$  & $\times$  & $\square$ & $\Diamond$ \\
ECoG               & $\bigcirc$ & + & $*$  & $\times$  & $\square$ & $\Diamond$ \\
Endovascular       & $\bigcirc$ & + &    & $\times$  & $\square$ &   \\ \bottomrule
\end{tabular}
\end{table}

\subsection{Amplitude and Signal-to-Noise Ratio (SNR)}
\begin{figure}[thb]
    \centering
    \begin{subfigure}[b]{0.49\textwidth}
        \includegraphics[trim={1.9cm 0 2.5cm 0},clip, width=\textwidth]{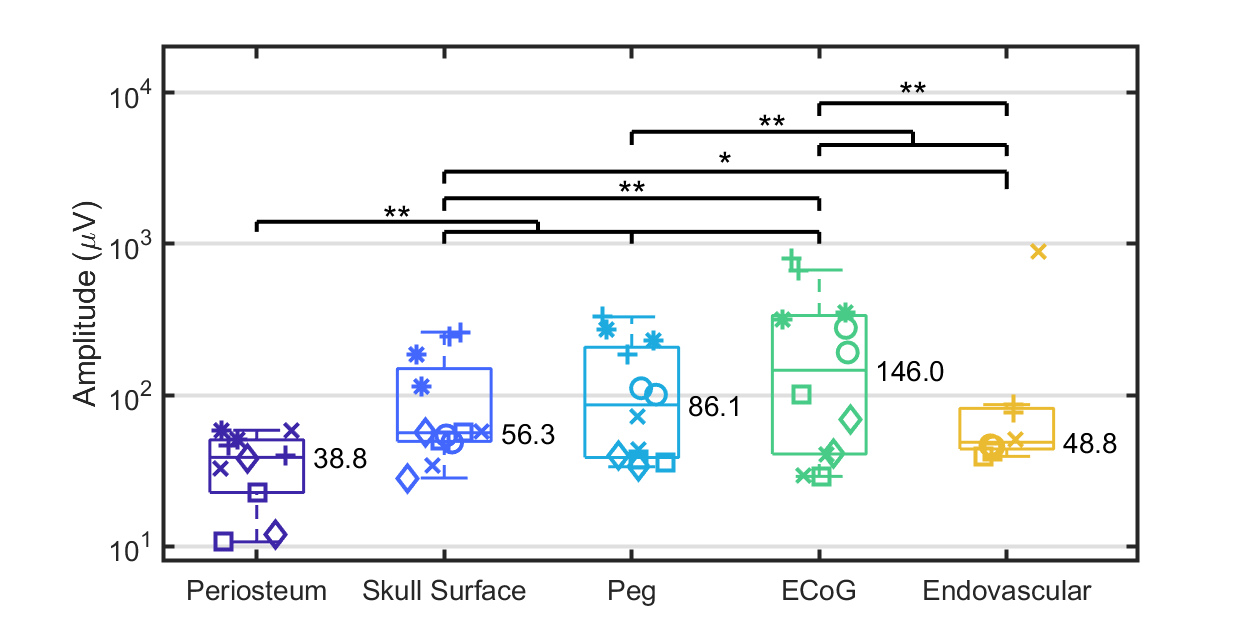}
        \caption{}
        \label{fig:CH3_Fig4_a}
    \end{subfigure}  
    \hfill
    \begin{subfigure}[b]{0.49\textwidth}
        \includegraphics[trim={1.9cm 0 2.5cm 0},clip,width=\textwidth]{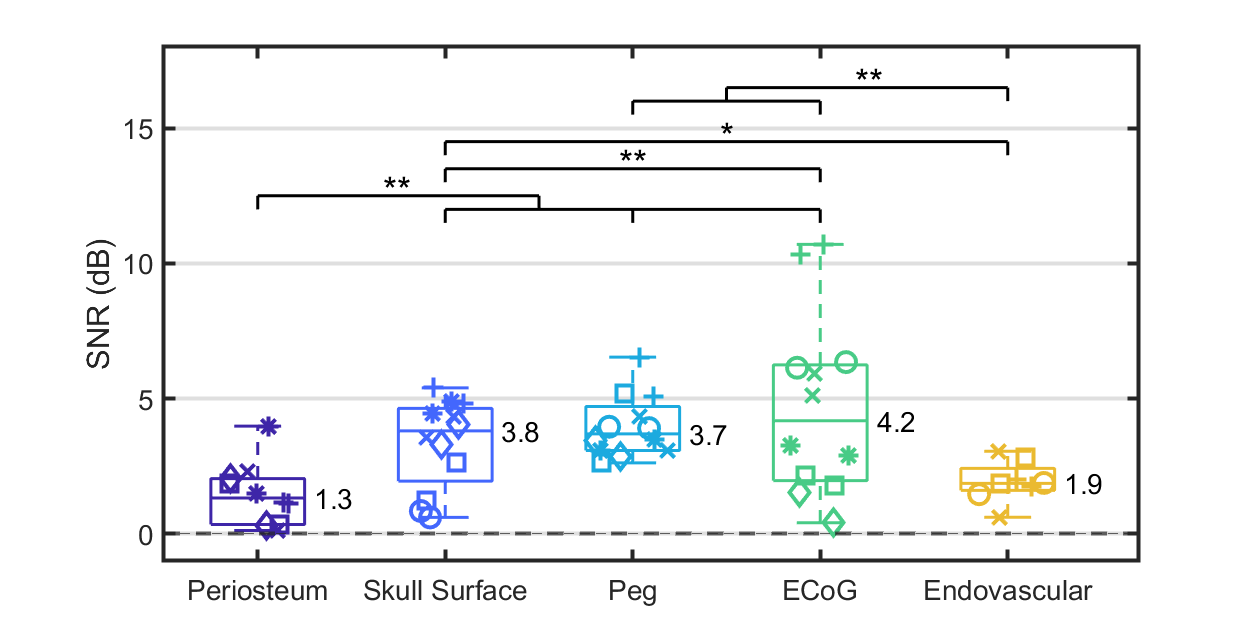}
        \caption{}
        \label{fig:CH3_Fig4_b}
    \end{subfigure} 
    \caption[VEP Amplitude and SNR]{Visual evoked potentials (VEPs) and signal-to-noise ratios (SNRs). (a) VEP amplitudes for different electrode depths. Each marker is the mean amplitude of a channel; the marker shapes indicate different sheep. The two channels with the highest SNR were used for each sheep. Median amplitudes are displayed beside each box. (b) The VEP SNR of each electrode location across sheep. Median VEP for each type is displayed beside each box. Asterisks indicate significant differences in the means ($*$~p$\leq$0.01, $**$~p$\leq$0.001, Tukey's HSD).}
\end{figure}

Highest median amplitude was recorded by ECoG electrodes (146~µV), followed by peg (86~µV), skull surface (56~µV), endovascular (49~µV), and periosteum (39~µV), as shown in Figure \ref{fig:CH3_Fig4_a}. A one-way analysis of variance (ANOVA) revealed a significant main effect from electrode location (F(4, 2595)=136.89, p$<$0.001) on VEP amplitude. Post-hoc pairwise Tukey's HSD tests, summarised in Table \ref{tab:CH3_Tab3}, indicated that VEP amplitudes recorded from ECoG electrodes were significantly greater than all other electrode locations (p$<$0.001). VEP amplitudes recorded from the periosteum were significantly lower than skull surface (p$<$0.001, Tukey's HSD) and peg (p$<$0.001, Tukey's HSD) electrodes, but were not significantly different to endovascular  electrodes (p=0.469, Tukey's HSD). VEP amplitudes recorded from the skull surface were not significantly different to peg electrodes (p=0.056, Tukey's HSD), but were significantly greater than endovascular electrodes (p=0.002, Tukey's HSD). 

All recording locations exhibited above 0 dB SNR, indicating VEPs with amplitude above noise were evoked in response to the stimuli, as shown in Figure \ref{fig:CH3_Fig4_b}. In line with the amplitude analyses, a one-way ANOVA revealed a significant main effect from electrode location on VEP SNR (F(4, 2595)=30.41, p$<$0.001). Across electrode locations, the greatest SNR was recorded by ECoG electrodes (median=4.2 dB), followed by skull surface (3.8 dB) and peg (3.7 dB), endovascular (1.9 dB), and periosteum electrodes (1.3 dB). Post-hoc testing with Tukey's HSD revealed periosteum electrodes recorded with significantly lower SNR than skull surface (p$<$0.001), peg (p$<$0.001), and ECoG electrodes (p$<$0.001), but with no significant difference to endovascular electrodes (p=0.74). Skull surface electrodes recorded with significantly lower SNR than ECoG (p$<$0.001, Tukey's HSD) and higher SNR than endovascular electrodes (p=0.003, Tukey's HSD), but not significantly lower than peg electrodes (p=0.29, Tukey's HSD). There was no significant difference in SNR between recordings from peg and ECoG electrodes (p=0.13, Tukey's HSD).

\begin{table}[ht]
\centering
\footnotesize
\caption[Pairwise Analysis of VEP Amplitude and SNR]{Results from post hoc pairwise Tukey's HSD comparison of VEP amplitude and SNR between each electrode location (significance $\alpha =$ 0.01).}
\label{tab:CH3_Tab3}
\begin{tabular}{llrrcrrc}
\toprule
                            \multicolumn{2}{c}{Electrode Location}                             & \multicolumn{3}{c}{Amplitude ($\mu$V)}                                                                          & \multicolumn{3}{c}{SNR (dB)}                                                                            \\\cmidrule(l){1-2} \cmidrule(l){3-5} \cmidrule(l){6-8}
\multicolumn{1}{c}{Group A} & \multicolumn{1}{c}{Group B} & \multicolumn{1}{c}{A-B} & \multicolumn{1}{c}{p} & \multicolumn{1}{c}{Significant} & \multicolumn{1}{c}{A-B} & \multicolumn{1}{c}{p} & \multicolumn{1}{c}{Significant} \\ \midrule
Periosteum                  & Surface                     & -62.00                           & $<$0.001      & Yes                                         & -1.87                        & $<$0.001      & Yes                                         \\
Periosteum                  & Peg                         & -86.93                           & $<$0.001      & Yes                                         & -2.49                        & $<$0.001      & Yes                                         \\
Periosteum                  & ECoG                        & -206.64                          & $<$0.001      & Yes                                         & -3.24                        & $<$0.001      & Yes                                         \\
Periosteum                  & Endovascular                & -19.26                           & 0.469                 & No                                          & -0.49                        & 0.736                 & No                                          \\
Surface                     & Peg                         & -24.92                           & 0.056                 & No                                          & -0.62                        & 0.288                 & No                                          \\
Surface                     & ECoG                        & -144.64                          & $<$0.001      & Yes                                         & -1.37                        & $<$0.001      & Yes                                         \\
Surface                     & Endovascular                & 42.74                            & 0.002                 & Yes                                         & 1.38                         & 0.003                 & Yes                                         \\
Peg                         & ECoG                        & -119.71                          & $<$0.001      & Yes                                         & -0.75                        & 0.1261                & No                                          \\
Peg                         & Endovascular                & 67.67                            & $<$0.001      & Yes                                         & 2.00                         & $<$0.001      & Yes                                         \\
ECoG                        & Endovascular                & 187.38                           & $<$0.001      & Yes                                         & 2.75                         & $<$0.001      & Yes                                         \\ \bottomrule
\end{tabular}
\end{table}

\subsection{Bandwidth}
\begin{figure}[ht]
    \centering
    \begin{subfigure}[b]{0.6\textwidth}
        \includegraphics[width=\textwidth]{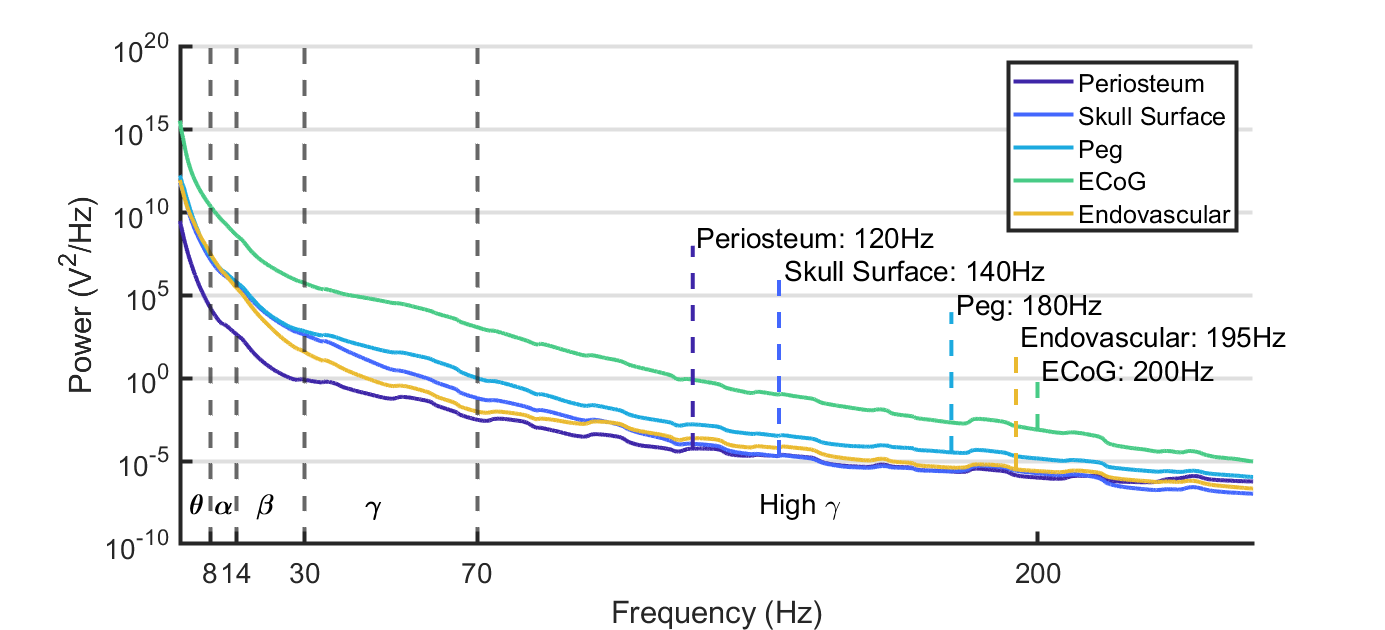}
        \caption{}
        \label{fig:CH3_Fig5_a}
    \end{subfigure}  
    \hfill
    \begin{subfigure}[b]{0.39\textwidth}
        \includegraphics[width=\textwidth]{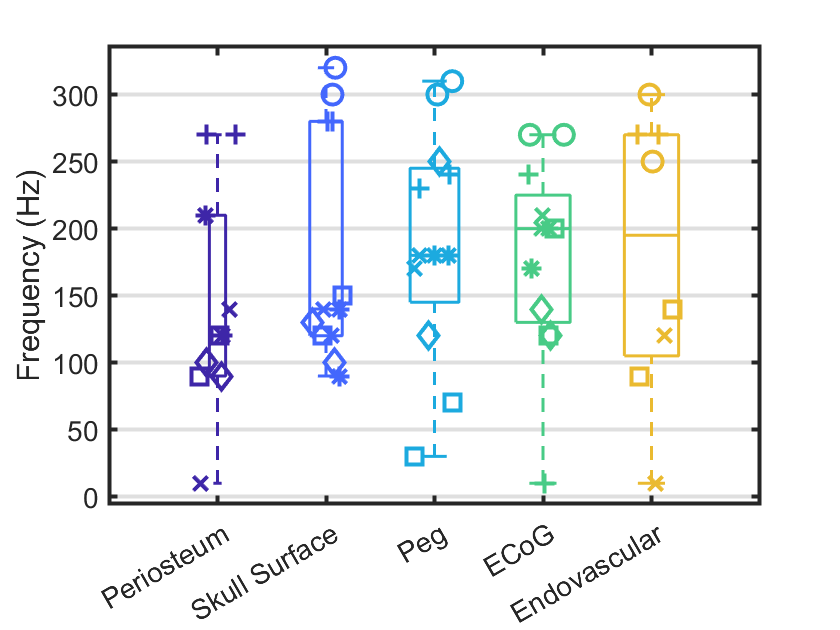}
        \caption{}
        \label{fig:CH3_Fig5_b}
    \end{subfigure} 
    \caption[Power Spectrum and Maximum Bandwidth]{Power spectra and maximum bandwidth. (a) Averaged power spectra across channels and sheep. Power across the spectrum increased with closer proximity to the brain. The median maximum bandwidth for each electrode location is indicated by a coloured dashed line. EEG bands are indicated by the black dashed lines. (b) Maximum bandwidths at each location. Each data point represents a channel; the marker styles indicate different animals.}
\end{figure}

Figure \ref{fig:CH3_Fig5_a} illustrates the power spectra of baseline EEG recordings from the electrode locations. ECoG exhibited the highest median maximum bandwidth (median 200~Hz) followed by endovascular (195~Hz), peg (180~Hz), skull surface (140~Hz) and periosteum (120~Hz), as shown in Figure \ref{fig:CH3_Fig5_b}. However, a one-way ANOVA revealed electrode location had no significant main effect on maximum bandwidth (F(4, 49)=0.47, p=0.76). Median maximum bandwidth for each electrode location was above 70~Hz, indicating all were recording high gamma neural activity. 

\subsection{Simulation of EEG signals}

\begin{figure}[ht]
    \centering
    \hfill
    \begin{subfigure}{0.45\linewidth}
        \includegraphics[trim={0.75cm -1.65cm 1.2cm 1.3cm},clip,width=\linewidth]{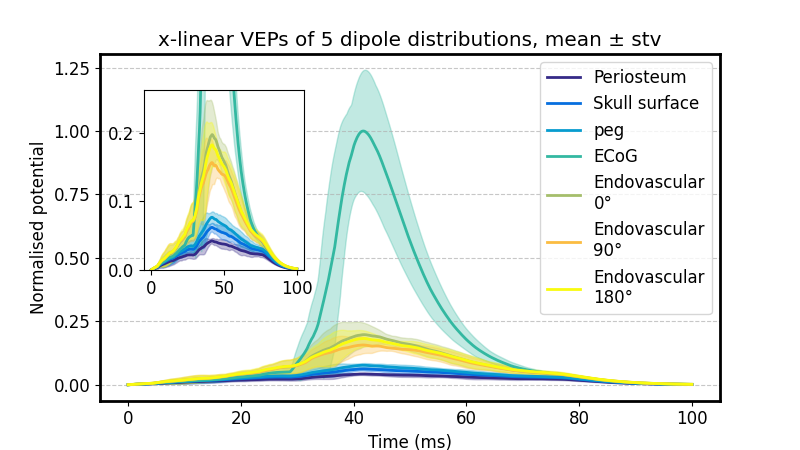}
        \caption{}
        \label{sim_VEPs}
    \end{subfigure}
    \hfill
    \begin{subfigure}{0.49\linewidth}
        \includegraphics[trim={0.75cm 0.2cm 1.5cm 0},clip,width=\linewidth]{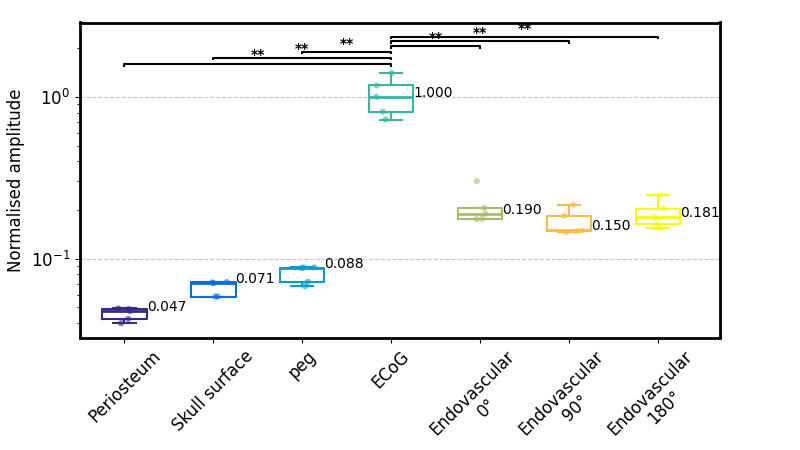}
        \caption{}
        \label{sim_boxplot}
    \end{subfigure}
    \caption{Simulated recordings of cortical \gls{epsp} with spatially dependent delays. (a) Time courses of simulated recordings for different electrodes. Solid trace shows the mean waveform of five random dipole distributions while the shaded area indicates a confidence level of one standard deviation. Inset shows the waveforms of the electrodes other than ECoG. (b) Amplitudes of simulated waveforms. Each marker represents the peak-to-peak amplitude of a simulated recording generated from one random dipole distribution. Median amplitudes are displayed beside each box. All amplitudes in (a) and (b) and have been normalised to the maximum mean and median amplitude, respectively. Asterisks indicate significant differences in the means ($**$~p$\leq$0.001, Tukey's HSD).}
    \label{sim_amplitudes}
\end{figure}

Seven electrode configurations were included in the simulations, namely, the periosteum, skull surface, peg within the skull, ECoG electrodes, and endovascular electrodes with three orientations relative to the cortex. Figure \ref{sim_amplitudes} shows the simulated waveforms and peak-to-peak amplitudes from different electrodes. The ECoG electrode yielded the highest amplitude compared to all other configurations (p$<$0.001, Tukey's HSD). The simulated amplitudes recorded by the endovascular electrodes were higher than those from the periosteum, skull surface, and peg electrodes; however, the differences were not significant (p = 0.232-0.850, Tukey's HSD). The three subscalp electrode depths and the orientation of the electrode electrodes did not result in any significant differences in simulated amplitudes (p$>$0.998 and p$>$0.996, respectively. Tukey's HSD). The simulated waveform from the ECoG electrode was the most variable with different dipole distributions, followed by the endovascular electrodes, while the three sub-scalp electrodes were the least affected.

\begin{figure}[htb]
    \centering
    \hfill
    \begin{subfigure}{0.49\linewidth}
        \includegraphics[trim={0.6cm 0.2cm 1.5cm 0},clip,width=\linewidth]{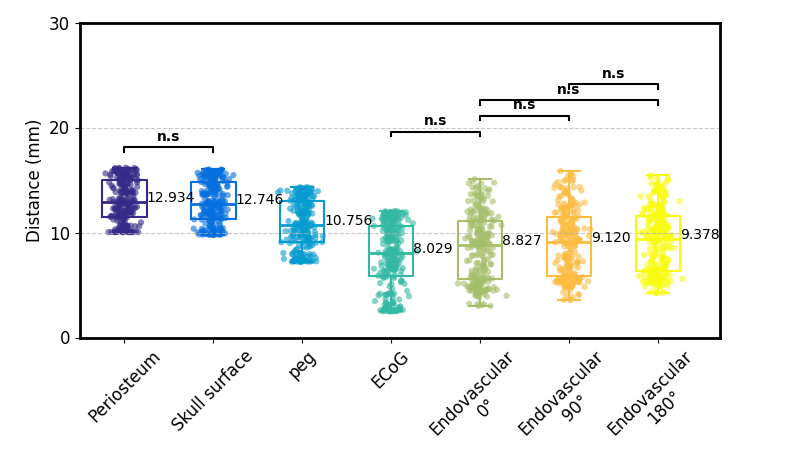}
        \caption{}
        \label{distance_boxplot}
    \end{subfigure}
    \hfill
    \begin{subfigure}{0.49\linewidth}
        \includegraphics[trim = {0.3cm 0.2cm 1.5cm 0},clip,width=\linewidth]{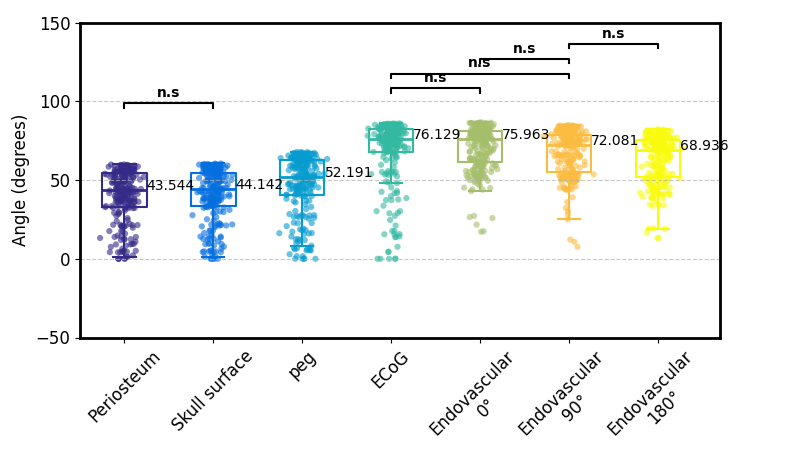}
        \caption{}
        \label{angle_boxplot}
    \end{subfigure}
    \vspace{0.5em}
    \hfill
    \begin{subfigure}{0.49\linewidth}
        \includegraphics[trim={0.75cm 0.05cm 1.5cm 1.3cm},clip,width=\linewidth]{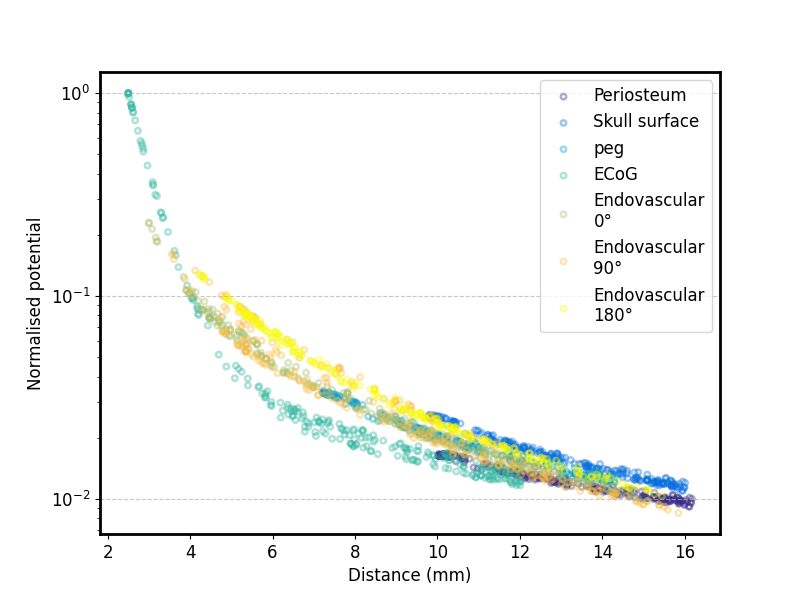}
        \caption{}
        \label{p2p_vs_distance}
    \end{subfigure}
    \hfill
    \begin{subfigure}{0.49\linewidth}
        \includegraphics[trim={0.75cm 0.05cm 1.5cm 1.3cm},clip,width=\linewidth]{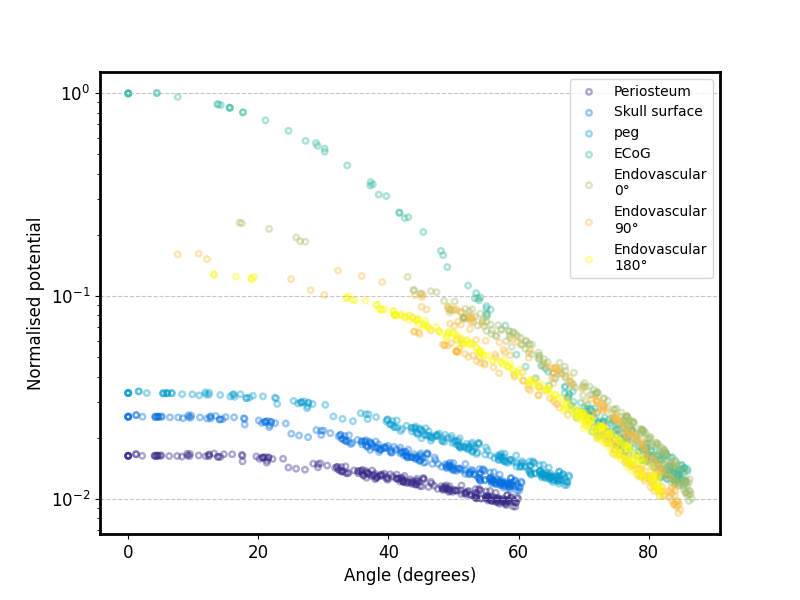}
        \caption{}
        \label{p2p_vs_angle}
    \end{subfigure}
    
    \caption{Effect of dipole-electrode distance and angle on static recorded potential. (a) Dipole-electrode distances of different electrode types. (b) Dipole-electrode angle of different electrode types. Median distances or angles are shown next to each box. Pairs marked with n.s. indicate no significant difference in the means (p$>$0.01, Tukey's HSD). (c) Static recorded potential with increasing dipole-electrode distance, and (d) with increasing dipole-electrode angle.All amplitudes are normalised relative to the maximum value. Each marker represents one simulated dipole position.}
    \label{distance_results}
\end{figure}

Figure \ref{distance_results} shows distances and relative angles between simulated dipoles and each recording electrode and their impacts on the static recorded potential. Dipole-electrode distances ranged from a median value of 8.03~mm with the ECoG electrode to 12.9~mm with the periosteum electrode. The ECoG electrode was significantly closer to the cortical dipoles than all other configurations (p$<$ 0.001, Tukey's HSD), except for the endovascular electrode directly facing the cortex (p=0.098). The peg electrode was significantly closer to the cortical dipoles compared to the other two subscalp depths, which also led to a significantly larger mean angle between the dipole moment and the vector pointing from the dipole to the electrode (p$<$0.001 for both comparisons). The subdural and endovascular electrodes also incurred significantly higher dipole-electrode relative angles compared to the subdural electrodes (p$<$0.001).
Static recorded potentials declined with the dipole-electrode distance, approximately following an inverse square law trend (Figure \ref{p2p_vs_distance}. Notably, potential recorded by the ECoG electrode declined more rapidly with distance than the other electrodes. Although the dipole-electrode distance of the endovascular electrode directly oriented towards the cortex was not significantly different from that of the ECoG electrode, the rate of decline in the recorded potential over distance was notably lower. At dipole-electrode distances greater than 10~mm, the skull surface electrode recorded higher potentials than the endovascular electrodes, while the periosteum electrode recorded lower potentials than the endovascular electrodes. Static recorded potentials also showed a gradual decline with dipole-electrode angle (Figure \ref{p2p_vs_angle}). 

\section{Discussion}
Sub-scalp EEG addresses limitations of current BCI signal acquisition technologies by being less invasive than intracranial electrodes, having improved stability and eliminating the tedious setup requirements of scalp EEG, and being spatially versatile and removable unlike endovascular electrodes. However, signal quality of activity recorded from the various layers in the sub-scalp space, particularly for signals evoked during BCI applications, is unknown. Here, we investigated the impact of sub-scalp electrode depth on signal quality by quantifying VEP SNR and maximum bandwidths for electrodes placed on the periosteum, directly on the skull, or partially within the skull. These results provide valuable insights for sub-scalp EEG BCI technology, especially regarding trade-offs between electrode location and signal quality. Furthermore, for the first time, we compared these characteristics with current gold standard BCI recording modalities: ECoG and endovascular electrode arrays.  

\subsection{Sub-scalp EEG Depths}
Through analysis of VEP SNR and maximum bandwidth, we quantitatively determined the signal quality afforded by placement of electrodes at three depths in the sub-scalp space: above the periosteum, directly on the surface of the skull, and embedded within the skull.

\subsubsection{Signal-to-Noise Ratio}
VEP responses varied between sub-scalp electrode depths. We observed an increase in SNR as the electrodes were placed closer to the brain. This is expected as signal amplitude attenuates with distance from the source. Electrodes placed on the periosteum recorded VEPs with only around 30\% of the SNR recorded by peg electrodes. These results are consistent with prior work by \citet{benovitski_ring_2017}, who concluded that peg electrodes saw significantly reduced artefact and higher EEG amplitude than disk or ring electrodes in sub-scalp space. Removing the periosteum and placing the electrode directly on the skull surface more than doubled the median VEP SNR. This result suggests a quick, simple method of electrode insertion to the sub-periosteal space may be a preferred solution, providing high quality signal via a minimally traumatic insertion procedure.

Embedding electrodes within the skull did not significantly increase VEP SNR over electrodes on the surface. The various risks, costs, and effort associated with the implantation of peg electrodes may outweigh the benefits in signal quality, particularly if the electrodes are to be burred into multiple sites across the skull to record activity from distal brain regions. 

\subsubsection{Bandwidth}
We saw an increase in median maximum bandwidth as the amount of tissue between the source and electrode reduced (periosteum: 120~Hz, skull surface: 140~Hz; peg: 180~Hz). All locations were able to detect high gamma band signal ($>$70~Hz). This result is consistent with previous work by \citet{olson_comparison_2016}, who demonstrated high gamma recordings in sub-scalp EEG. Numerous studies have demonstrated high gamma band recordings with traditional surface EEG setups in humans \citep{smith_non-invasive_2014, ball_movement_2008, guger_utilizing_2013}, so it is reasonable to expect that sub-scalp EEG can also record high-gamma activity. The similarity between these results and previous studies supports the validity of methods adopted in this study and the reported maximum bandwidths for sub-scalp electrodes. High gamma band power has been shown to be modulated in relation to motor function and can be volitionally modulated as a means of BCI control \citep{smith_non-invasive_2014, ball_movement_2008, guger_utilizing_2013}. As such, by demonstrating high gamma band activity from sub-scalp EEG recordings, these results support the signal acquisition method for BCI applications. As high gamma activity was recorded above the periosteal layer, simply tunnelling electrodes beneath the scalp may allow for decoding of high gamma features useful for BCI applications although deeper layers, such as skull surface and peg electrodes, may provide more fidelity.

\subsection{Sub-scalp EEG Comparison with Gold Standard Modalities}
In addition to providing insight into signal quality of sub-scalp EEG depths, the results of this study have characterised sub-scalp EEG in the broader neural signal acquisition space by way of comparison with the current gold standard modalities: ECoG and endovascular arrays.

\subsubsection{ECoG}

ECoG, as expected due to its proximity to the activity source, demonstrated the highest signal quality across all modalities. Periosteum and skull surface electrodes recorded VEP SNR significantly lower than ECoG SNR; however, peg electrodes were not significantly different from ECoG. Peg electrodes also approached ECoG signal quality in terms of maximum bandwidth (peg: 180~Hz, ECoG: 200~Hz). These results indicate that peg electrodes may provide a less invasive alternative to ECoG arrays. 

However, the ECoG recording performance may have been reduced by limitations in the experimental setup. The ECoG recordings were performed approximately 6-8 hours after stent and periosteum recordings (which were performed first), due to being the deepest of the methods inserted through the scalp. In sheep exhibiting low ECoG VEP amplitudes, it is possible that eye deterioration due to dryness resulted in reduced VEP responses, despite regular application of saline solution to maintain eye health. Alternatively, the prolonged delivery of anaesthesia may have affected VEP responses in some sheep more than others. Future work may benefit from changing the order of recording methods, performing the recordings simultaneously, or performing the experiment in awake subjects to mitigate these limitations. 

\subsubsection{Endovascular}

There was no significant difference between VEP SNR recorded by endovascular electrodes (median 1.9~dB) and electrodes on the periosteum (1.3~dB). Regarding the frequency space, endovascular electrodes demonstrated relatively high median maximum bandwidth (195~Hz). While we expect a minor improvement to endovascular electrode performance with chronic implantation due to endothelialisation (incorporation into the blood vessel wall) \citep{opie_micro-ct_2017}, the comparable SNR demonstrated by the modality in comparison to less invasive sub-scalp electrodes suggests sub-scalp EEG is the more suitable signal acquisition method. 

The stent implantation procedure requires a highly skilled surgeon and the complexity in the procedure will only increase as attempts are made to broaden the spatial capabilities of the technology by reaching into sub-branching vessels. Reaching these vessels is critical for the future development of endovascular arrays due to current limitations in spatial capabilities, being only able to detect activity from brain regions proximal to major vessels such as the superior sagittal sinus. With increased procedure complexity comes increased risk and cost to the patient. Sub-scalp electrodes, in comparison, can be trivially placed over brain regions of interest, and even span several regions through a single small incision via tunnelling \citep{haneef_sub-scalp_2022}. Additionally, sub-scalp EEG mitigates the risks associated with stent implantation, such as thrombosis \citep{starke_endovascular_2015, modi_stent_2024}, and can be removed easily \citep{benovitski_ring_2017}.

Within the setup and parameters tested in this study, sub-scalp EEG did not achieve the same signal quality as ECoG arrays, which is expected. However, it may provide a safer and simpler alternative to endovascular arrays with comparable signal quality and improved spatial capabilities.

\subsection{Forward Simulation of EEG signals}
The tissue volume conductor model, assuming purely resistive tissue dielectric conductivities, predicted that the ECoG electrode recorded the highest amplitude from the equivalent cortical dipoles, whereas the recording amplitudes of the sub-scalp electrodes were not significantly different from those of the endovascular electrodes. This prediction aligned well with the experimental observations, suggesting that frequency-invariant tissue conductivity and dipole-electrode distance alone account for a large portion of the discrepancy in the recorded VEP amplitudes across electrode depths. The orientation of the endovascular electrodes did not result in a significant difference in recording amplitude, which is consistent with a previous \textit{in vivo} study \citep{opie_effect_2018}.  

However, although not significant, the model predicted overall higher recording amplitudes from the endovascular electrodes compared to the skull surface electrode, contrary to the experimental observations. This discrepancy may largely stem from the differences in electrode-tissue interface impedance due to variations in electrode surface areas, which were not considered in the simulations. Endovascular electrodes (0.75~mm diameter) have only 6.25\% of the surface area of that of the skull surface electrodes (3~mm diameter), which in practice would notably impact their performance and reduce signal amplitudes. Similarly, the model may have overestimated the performance of the ECoG electrode (1.5~mm diameter), as its surface area is four times smaller than that of the skull surface electrode and eight times smaller than that of the peg electrode.

The discrepancies in the simulated recorded static potential of each dipole across different electrode depths resulted from tissue conductivity, dipole-electrode distance, and dipole orientation relative to the electrode. In this simplified brain model, the cortex was assumed to be smooth, with only radial dipoles included. This implies that electrodes positioned further away from the cortex also had smaller dipole-electrode angles. Electrodes are more polarised by dipoles with smaller relative angles than those with an angle approaching 90$^\circ$. The effect of dipole orientation may explain the reversed trend in static recorded potential amplitude at large dipole-electrode distances above 10~mm, where the skull surface electrode recorded the highest amplitudes and ECoG the lowest. In contrast, the periosteum electrode recorded lower amplitudes than the endovascular electrodes despite the smaller dipole-electrode angles, highlighting the high impedance of the periosteum. Moreover, this modeling observation underscores dipole orientation as a confounding factor, implying preferential recording of tangential superficial dipoles by electrodes close to the cortex and of radial superficial dipoles by electrodes placed further away.

Overall, the endovascular electrodes, despite having similar distances and relative angles to the cortical dipoles as the ECoG electrode, yielded significantly lower recording amplitudes in the simulated waveforms. The amplitudes recorded by sub-scalp electrodes were not significantly different from those of the endovascular electrodes. The simulated EEG recordings, in conjunction with the experimental measurements, highlighted the suitability of sub-scalp EEG as a less invasive alternative to the endovascular approach.

\subsection{The Future of Sub-scalp BCI}
\subsubsection{Large Data Models}
The advent of continuous EEG streaming from a potentially large cohort of users provides the opportunity for several interesting research avenues, both within and outside the context of BCI. Large banks of sub-scalp BCI data may permit the training of sophisticated AI algorithms, allowing these minimally invasive systems to approach levels of functional performance previously only presumed to be possible with ECoG or penetrating electrodes. Data gathered across large cohorts of users could be used to pre-train BCI classifiers, reducing the amount of data required to gather from new users to calibrate their own personal classifiers, thereby reducing training time for users from months or weeks to days or even hours. 

\subsubsection{Sub-Scalp BCI Performance and Patient Cohorts}
Since the results suggest similar SNR between endovascular recording and sub-scalp EEG, we expect a comparable performance in BCI applications. Currently, endovascular arrays have been chronically implanted in human patients. These patients have demonstrated consistent ability to produce binary output, providing a `click' during computer-assisted scrolling \citep{oxley_motor_2021}. This output has a relatively low dimensionality when compared to previous invasive BCI implementations controlling robotic arms \citep{hochberg_reach_2012} and non-invasive systems controlling a three-dimensional cursor \citep{mcfarland_electroencephalographic_2010}, and may be due to the limited spatial capabilities of endovascular arrays. As sub-scalp EEG can record from multiple regions of the brain, we expect sub-scalp BCIs to offer improved dimensionality over endovascular arrays. However, spatial resolution of electrodes implanted in the sub-scalp space has not been investigated. 

With non-invasive EEG providing a benchmark, prior literature suggests that a user with a sub-scalp EEG BCI could achieve simultaneous control of three dimensions \citep{mcfarland_electroencephalographic_2010}. This level of control would allow users to operate several devices, including prostheses, mobility aids, keyboards, and perhaps most importantly, computer cursor/click systems. Simple cursor control provides individuals with disability access to a wide range of innovative services and applications available through the internet and Internet of Things, including those that offer everyday assistance, support, social connection, leisure, entertainment, employment, and so on. Highly invasive techniques, such as intracranial electrodes, focus on providing high-dimensional, sophisticated BCI control. The benefit provided by intracranial devices may outweigh the risks in cases where the user experiences significant impairment in motor functions, such as those with MND or brain-stem stroke, given the BCI may significantly improve the user's quality of life, and considering that life expectancy for these individuals is often limited. However, intracranial BCI devices may not be appropriate for individuals with less severe conditions and longer life expectancy due to the high level of risk associated with these implants. The minimally invasive nature of sub-scalp EEG suggests it is more appropriate for much larger cohorts who may have trouble with more general hand motor function, such as those with multiple sclerosis, cerebral palsy, spinal cord injury, amputation, muscular dystrophy, or people undergoing stroke rehabilitation. While sub-scalp EEG may not provide the same level of control as more invasive electrodes, it may provide sufficient control (through cursor function) to improve quality of life for these cohorts with much lower risk. Having such a wider potential user group also improves commercialisation prospects for this technology. For these reasons, there is promise that sub-scalp BCI will become widely prevalent in future as BCIs become more commercially available and as sub-scalp EEG continues to demonstrate safety and benefits through other clinical applications, such as seizure monitoring \citep{duun-henriksen_new_2020}. 

\subsubsection{Additional Chronic EEG Indications}
In addition to BCI, high throughput sub-scalp EEG could also monitor other conditions commonly experienced by target users, such as pain \citep{zis_eeg_2022}, fatigue \citep{qi_neural_2019}, stress \citep{perez-valero_quantitative_2021}, depression \citep{neto_depression_2019}, apnea \citep{zhao_classification_2021}, and so on. Providing this information to clinicians may result in better health outcomes for users. In some cases, monitoring these conditions may provide enough benefit to support implantation of sub-scalp EEG devices in persons without severe paralysis, as in the case of long-term seizure monitoring for persons with epilepsy \citep{stirling_seizure_2021}. 

\subsection{Limitations}
This study includes several limitations that should be addressed in future work. 

\subsubsection{Acute Experiment}
The acute nature of the experiment may produce results that vary from chronically implanted electrodes. Endovascular arrays undergo endothelialisation into the vessel wall within weeks of implantation, improving signal quality \citep{opie_micro-ct_2017}. Additionally, 10-week implantation has shown tissue growth around ring electrodes, which may result in poorer EEG signal, and disk electrodes on the surface of the skull tend to erode into the bone, which may improve EEG signal \citep{benovitski_ring_2017}. While acute experiments may provide some insight into electrode performance, future work should aim to assess BCI control after chronic implantation, as would be expected of a chronically implanted BCI system.

\subsubsection{Anaesthesia}
EEG is affected by anaesthesia. Notably, isoflurane has been shown to attenuate high gamma band EEG \citep{plourde_attenuation_2016}. Therefore, maximum bandwidth in awake subjects is expected to be increased. However, as the use of anaesthesia was consistent across animals and electrode locations, the relative differences in signal quality observed between electrode locations are expected to be similar for awake sheep models.

\subsubsection{Impedance}
Surface area was not consistent between the different electrode types, influencing electrode impedance, spatial resolution, and signal quality. This was difficult to control due to the various types of arrays used in the experiment. Measuring electrode impedance may have provided insight into how the surface areas affected electrode performance; however, this was not performed in this study.

\subsubsection{Spatial Resolution}
It should be noted that SNR and bandwidth are not the only measures of signal quality. Spatial resolution is another important metric that can impact BCI performance. High spatial resolution of ECoG recordings allows for discrimination between local sources of brain activity, allowing for more accurate and higher dimensional control compared to scalp EEG. Peg electrodes, which approach the depth of ECoG arrays, may provide sufficient spatial resolution to produce BCI functionality comparable to those previously demonstrated with ECoG \citep{benabid_exoskeleton_2019}. Such functionality may warrant their more invasive implantation procedure. Investigation into sub-scalp EEG spatial resolution will provide further insight toward its fitness for BCI applications, as well as inform design choices that were not addressed in this study, such as optimal electrode size and pitch.

\subsubsection{Animal Models}
Sheep were used in this study due to their similar skull size and vasculature to human models. However, variations in anatomy, such as skull thickness and neural responses, may limit how well these results translate to humans. The sheep were also under general anaesthesia during the experiment. Anaesthesia can affect the brain's response to stimulus. Variations in anaesthesia between animals, and within animals between electrode type recordings, may affect the neural activity recorded. Additionally, the neural activity was not polluted by artefacts one might expect during EEG recording in the home, such as EMG, EOG, or other common neural processes in an awake human. The various recording methods tested in this study may be affected differently by these limitations. As such, future work should focus on investigating sub-scalp BCI in awake animals or advancing to in-human research.

\subsubsection{Simulation Assumptions}
The tissue conductor model used in this study was a simplification of the electrical properties of biological tissues. Several assumptions and limitations must be considered when interpreting the results: 1) The tissues were assumed to be purely resistive with frequency-independent conductivities. Frequency-dependent signal dispersion at tissue interfaces was not considered, and signal attenuation was attributed solely to ohmic losses. This assumption becomes less valid for higher-frequency signals, and the low-pass filter properties of the skull, which can lead to dispersive signal distortion, were not accounted for. 2) Dielectric properties and capacitive effects of tissues were not included in the simulations. As a result, charge buildup at heterogeneous tissue interfaces was not considered, which affects the temporal evolution of the electric field and thus impacts the recorded waveforms. 3) Tissues were assumed to have isotropic conductivities, which is not realistic and would affect the trajectory of current flow. 4) The impedance and dielectric properties of the electrode-tissue interface were not considered, though these are expected to have a significant impact on the recorded waveforms. 5) Cortex geometry and the spatial dependency of EPSP propagation were simplified. 6) Physiological noise was not considered, which may have varying effects at different electrode depths.

\section{Conclusion}
Sub-scalp EEG addresses limitations of current BCI signal acquisition methods by providing a low risk alternative to ECoG and endovascular arrays, with improved stability and aesthetic appeal over surface EEG. Understanding how the tissue layers in the sub-scalp space impact signal quality will inform future sub-scalp BCI implant designs. Here, we quantified signal quality of EEG recorded from different depths in the sub-scalp space and benchmarked their quality with current gold standard BCI recording methods: ECoG and endovascular arrays. Key insights include that, in response to visual stimuli, peg electrode VEP SNR approaches SNR of ECoG, and endovascular arrays have SNR comparable to electrodes placed on the periosteum. Furthermore, we demonstrated that sub-scalp electrodes capture high gamma neural activity, with maximum bandwidth ranging from 120~Hz to 180~Hz, depending on electrode depth. The outcomes of this study will guide future sub-scalp BCI device and electrode design choices, providing quantitative evidence to support the use of more invasive peg electrodes, in the case where high SNR or bandwidth is required, or simple tunnelled electrodes otherwise. More generally, these results support the use of sub-scalp EEG for BCI applications, propelling this technology toward chronic, in-home BCI applications that improve quality of life for those living with physical disability.

\section{Acknowledgements}
We thank research assistant Huakun Xin, veterinary technician Tomas Vale, and animal handler Quan Nguyen (The Florey Institute of Neuroscience and Mental Health) for their assistance in caring for the animals, experiment preparations, and monitoring during the animal experiments.
We also wish to thank the animals used for this research. Animal studies are crucial for the realisation of medical devices. We intend to ensure insights gained from their sacrifice lead to improved quality of life for persons with significant functional impairment.
FEM simulations were performed with Sim4Life by ZMT, www.zmt.swiss.


\begin{thebibliography}{56}
\providecommand{\natexlab}[1]{#1}
\providecommand{\url}[1]{\texttt{#1}}
\expandafter\ifx\csname urlstyle\endcsname\relax
  \providecommand{\doi}[1]{doi: #1}\else
  \providecommand{\doi}{doi: \begingroup \urlstyle{rm}\Url}\fi

\bibitem[Ball et~al.(2008)Ball, Demandt, Mutschler, Neitzel, Mehring, Vogt, Aertsen, and Schulze-Bonhage]{ball_movement_2008}
T.~Ball, E.~Demandt, I.~Mutschler, E.~Neitzel, C.~Mehring, K.~Vogt, A.~Aertsen, and A.~Schulze-Bonhage.
\newblock Movement related activity in the high gamma range of the human {EEG}.
\newblock \emph{NeuroImage}, 41\penalty0 (2):\penalty0 302--310, June 2008.
\newblock ISSN 10538119.
\newblock \doi{10.1016/j.neuroimage.2008.02.032}.
\newblock URL \url{https://linkinghub.elsevier.com/retrieve/pii/S1053811908001717}.

\bibitem[Ball et~al.(2009)Ball, Kern, Mutschler, Aertsen, and Schulze-Bonhage]{ball_signal_2009}
T.~Ball, M.~Kern, I.~Mutschler, A.~Aertsen, and A.~Schulze-Bonhage.
\newblock Signal quality of simultaneously recorded invasive and non-invasive {EEG}.
\newblock \emph{NeuroImage}, 46\penalty0 (3):\penalty0 708--716, 2009.
\newblock ISSN 1053-8119.
\newblock \doi{10.1016/j.neuroimage.2009.02.028}.
\newblock URL \url{https://www.sciencedirect.com/science/article/pii/S1053811909001827}.

\bibitem[Benabid et~al.(2019)Benabid, Costecalde, Eliseyev, Charvet, Verney, Karakas, Foerster, Lambert, Morinière, Abroug, Schaeffer, Moly, Sauter-Starace, Ratel, Moro, Torres-Martinez, Langar, Oddoux, Polosan, Pezzani, Auboiroux, Aksenova, Mestais, and Chabardes]{benabid_exoskeleton_2019}
A.~L. Benabid, T.~Costecalde, A.~Eliseyev, G.~Charvet, A.~Verney, S.~Karakas, M.~Foerster, A.~Lambert, B.~Morinière, N.~Abroug, M.-C. Schaeffer, A.~Moly, F.~Sauter-Starace, D.~Ratel, C.~Moro, N.~Torres-Martinez, L.~Langar, M.~Oddoux, M.~Polosan, S.~Pezzani, V.~Auboiroux, T.~Aksenova, C.~Mestais, and S.~Chabardes.
\newblock An exoskeleton controlled by an epidural wireless brain–machine interface in a tetraplegic patient: a proof-of-concept demonstration.
\newblock \emph{The Lancet Neurology}, 18\penalty0 (12):\penalty0 1112--1122, Dec. 2019.
\newblock ISSN 14744422.
\newblock \doi{10.1016/S1474-4422(19)30321-7}.
\newblock URL \url{https://linkinghub.elsevier.com/retrieve/pii/S1474442219303217}.

\bibitem[Benovitski et~al.(2017)Benovitski, Lai, McGowan, Burns, Maxim, Nayagam, Millard, Rathbone, le~Chevoir, Williams, Grayden, May, Murphy, D’Souza, Cook, and Williams]{benovitski_ring_2017}
Y.~B. Benovitski, A.~Lai, C.~C. McGowan, O.~Burns, V.~Maxim, D.~A.~X. Nayagam, R.~Millard, G.~D. Rathbone, M.~A. le~Chevoir, R.~A. Williams, D.~B. Grayden, C.~N. May, M.~Murphy, W.~J. D’Souza, M.~J. Cook, and C.~E. Williams.
\newblock Ring and peg electrodes for minimally-{Invasive} and long-term sub-scalp {EEG} recordings.
\newblock \emph{Epilepsy Research}, 135:\penalty0 29--37, 2017.
\newblock ISSN 0920-1211.
\newblock \doi{10.1016/j.eplepsyres.2017.06.003}.
\newblock Type: Journal Article.

\bibitem[Chaudhary et~al.(2016)Chaudhary, Birbaumer, and Ramos-Murguialday]{chaudhary_chapter_2016}
U.~Chaudhary, N.~Birbaumer, and A.~Ramos-Murguialday.
\newblock Chapter 5 - {Brain}-computer interfaces in the completely locked-in state and chronic stroke.
\newblock In D.~Coyle, editor, \emph{Progress in {Brain} {Research}}, volume 228, pages 131--161. Elsevier, 2016.
\newblock ISBN 0079-6123.
\newblock \doi{https://doi.org/10.1016/bs.pbr.2016.04.019}.
\newblock Type: Book Section.

\bibitem[Clarke and Whitteridge(1976)]{clarke_cortical_1976}
P.~G. Clarke and D.~Whitteridge.
\newblock The cortical visual areas of the sheep.
\newblock \emph{The Journal of Physiology}, 256\penalty0 (3):\penalty0 497--508, Apr. 1976.
\newblock ISSN 0022-3751, 1469-7793.
\newblock \doi{10.1113/jphysiol.1976.sp011335}.
\newblock URL \url{https://physoc.onlinelibrary.wiley.com/doi/10.1113/jphysiol.1976.sp011335}.

\bibitem[Duun-Henriksen et~al.(2020)Duun-Henriksen, Baud, Richardson, Cook, Kouvas, Heasman, Friedman, Peltola, Zibrandtsen, and Kjaer]{duun-henriksen_new_2020}
J.~Duun-Henriksen, M.~Baud, M.~P. Richardson, M.~Cook, G.~Kouvas, J.~M. Heasman, D.~Friedman, J.~Peltola, I.~C. Zibrandtsen, and T.~W. Kjaer.
\newblock A new era in electroencephalographic monitoring? {Subscalp} devices for ultra-long-term recordings.
\newblock \emph{EPILEPSIA}, 2020.
\newblock ISSN 00139580.
\newblock \doi{10.1111/epi.16630}.
\newblock Type: Journal Article.

\bibitem[Feigin et~al.(2020)Feigin, Vos, Nichols, Owolabi, Carroll, Dichgans, Deuschl, Parmar, Brainin, and Murray]{feigin_global_2020}
V.~L. Feigin, T.~Vos, E.~Nichols, M.~O. Owolabi, W.~M. Carroll, M.~Dichgans, G.~Deuschl, P.~Parmar, M.~Brainin, and C.~Murray.
\newblock The global burden of neurological disorders: translating evidence into policy.
\newblock \emph{The Lancet Neurology}, 19\penalty0 (3):\penalty0 255--265, Mar. 2020.
\newblock ISSN 14744422.
\newblock \doi{10.1016/S1474-4422(19)30411-9}.
\newblock URL \url{https://linkinghub.elsevier.com/retrieve/pii/S1474442219304119}.

\bibitem[Haneef et~al.(2022)Haneef, Yang, Sheth, Aloor, Aazhang, Krishnan, and Karakas]{haneef_sub-scalp_2022}
Z.~Haneef, K.~Yang, S.~A. Sheth, F.~Z. Aloor, B.~Aazhang, V.~Krishnan, and C.~Karakas.
\newblock Sub-scalp electroencephalography: {A} next-generation technique to study human neurophysiology.
\newblock \emph{Clinical Neurophysiology}, 141:\penalty0 77--87, 2022.
\newblock ISSN 1388-2457.
\newblock \doi{10.1016/j.clinph.2022.07.003}.
\newblock URL \url{https://www.sciencedirect.com/science/article/pii/S1388245722003273}.

\bibitem[Hochberg et~al.(2012)Hochberg, Bacher, Jarosiewicz, Masse, Simeral, Vogel, Haddadin, Liu, Cash, Van Der~Smagt, and Donoghue]{hochberg_reach_2012}
L.~R. Hochberg, D.~Bacher, B.~Jarosiewicz, N.~Y. Masse, J.~D. Simeral, J.~Vogel, S.~Haddadin, J.~Liu, S.~S. Cash, P.~Van Der~Smagt, and J.~P. Donoghue.
\newblock Reach and grasp by people with tetraplegia using a neurally controlled robotic arm.
\newblock \emph{Nature}, 485\penalty0 (7398):\penalty0 372--375, May 2012.
\newblock ISSN 0028-0836, 1476-4687.
\newblock \doi{10.1038/nature11076}.
\newblock URL \url{https://www.nature.com/articles/nature11076}.

\bibitem[{IT'IS Foundation}(2024)]{itis_foundation_tissue_2024}
{IT'IS Foundation}.
\newblock Tissue {Properties} {Database} {V4}.2, 2024.
\newblock URL \url{https://itis.swiss/virtual-population/tissue-properties/downloads/database-v4-2/}.

\bibitem[Jiang et~al.(2017)Jiang, Gu, Li, Gao, Sun, Song, Wang, Yuan, Wang, Liu, Han, and Dai]{jiang_analysis_2017}
Y.~Jiang, P.~Gu, B.~Li, X.~Gao, B.~Sun, Y.~Song, G.~Wang, Y.~Yuan, C.~Wang, M.~Liu, D.~Han, and P.~Dai.
\newblock Analysis and {Management} of {Complications} in a {Cohort} of 1,065 {Minimally} {Invasive} {Cochlear} {Implantations}.
\newblock \emph{Otology \& Neurotology}, 38\penalty0 (3):\penalty0 347--351, Mar. 2017.
\newblock ISSN 1531-7129, 1537-4505.
\newblock \doi{10.1097/MAO.0000000000001302}.
\newblock URL \url{https://journals.lww.com/00129492-201703000-00006}.

\bibitem[John et~al.(2018)John, Opie, Wong, Rind, Ronayne, Gerboni, Bauquier, O’Brien, May, Grayden, and Oxley]{john_signal_2018}
S.~E. John, N.~L. Opie, Y.~T. Wong, G.~S. Rind, S.~M. Ronayne, G.~Gerboni, S.~H. Bauquier, T.~J. O’Brien, C.~N. May, D.~B. Grayden, and T.~J. Oxley.
\newblock Signal quality of simultaneously recorded endovascular, subdural and epidural signals are comparable.
\newblock \emph{Scientific Reports}, 8\penalty0 (1):\penalty0 8427, 2018.
\newblock ISSN 2045-2322.
\newblock \doi{10.1038/s41598-018-26457-7}.
\newblock URL \url{https://doi.org/10.1038/s41598-018-26457-7}.
\newblock Type: Journal Article.

\bibitem[Johnston et~al.(2006)Johnston, Mangano, Ojemann, Park, Trevathan, and Smyth]{johnston_complications_2006}
J.~M. Johnston, F.~T. Mangano, J.~G. Ojemann, T.~S. Park, E.~Trevathan, and M.~D. Smyth.
\newblock Complications of invasive subdural electrode monitoring at {St}. {Louis} {Children}’s {Hospital}, 1994–2005.
\newblock \emph{Journal of Neurosurgery: Pediatrics PED}, 105\penalty0 (5):\penalty0 343--347, 2006.
\newblock \doi{10.3171/ped.2006.105.5.343}.
\newblock Type: Journal Article.

\bibitem[Khan et~al.(2020)Khan, Das, Iversen, and Puthusserypady]{khan_review_2020}
M.~A. Khan, R.~Das, H.~K. Iversen, and S.~Puthusserypady.
\newblock Review on motor imagery based {BCI} systems for upper limb post-stroke neurorehabilitation: {From} designing to application.
\newblock \emph{Computers in Biology and Medicine}, 123:\penalty0 103843, 2020.
\newblock ISSN 0010-4825.
\newblock \doi{10.1016/j.compbiomed.2020.103843}.
\newblock Type: Journal Article.

\bibitem[Kinaci et~al.(2020)Kinaci, Bergmann, Bleys, Van Der~Zwan, and Van~Doormaal]{kinaci_histologic_2020}
A.~Kinaci, W.~Bergmann, R.~L. Bleys, A.~Van Der~Zwan, and T.~P. Van~Doormaal.
\newblock Histologic {Comparison} of the {Dura} {Mater} among {Species}.
\newblock \emph{Comparative Medicine}, 70\penalty0 (2):\penalty0 170--175, Apr. 2020.
\newblock ISSN 1532-0820.
\newblock \doi{10.30802/AALAS-CM-19-000022}.
\newblock URL \url{https://aalas.kglmeridian.com/view/journals/72010023/70/2/article-p170.xml}.

\bibitem[Kubler et~al.(2014)Kubler, Holz, Riccio, Zickler, Kaufmann, Kleih, Staiger-Salzer, Desideri, Hoogerwerf, and Mattia]{kubler_user-centered_2014}
A.~Kubler, E.~M. Holz, A.~Riccio, C.~Zickler, T.~Kaufmann, S.~C. Kleih, P.~Staiger-Salzer, L.~Desideri, E.~J. Hoogerwerf, and D.~Mattia.
\newblock The user-centered design as novel perspective for evaluating the usability of {BCI}-controlled applications.
\newblock \emph{PLoS One}, 9\penalty0 (12):\penalty0 e112392, 2014.
\newblock ISSN 1932-6203 (Electronic) 1932-6203 (Linking).
\newblock \doi{10.1371/journal.pone.0112392}.
\newblock Type: Journal Article.

\bibitem[Logroscino et~al.(2018)Logroscino, Piccininni, Marin, Nichols, Abd-Allah, Abdelalim, Alahdab, Asgedom, Awasthi, Chaiah, Daryani, Do, Dubey, Elbaz, Eskandarieh, Farhadi, Farzadfar, Fereshtehnejad, Fernandes, Filip, Foreman, Gebre, Gnedovskaya, Hamidi, Hay, Irvani, Ji, Kasaeian, Kim, Mantovani, Mashamba-Thompson, Mehndiratta, Mokdad, Nagel, Nguyen, Nixon, Olagunju, Owolabi, Piradov, Qorbani, Radfar, Reiner, Sahraian, Sarvi, Sharif, Temsah, Tran, Truong, Venketasubramanian, Winkler, Yimer, Feigin, Vos, and Murray]{logroscino_global_2018}
G.~Logroscino, M.~Piccininni, B.~Marin, E.~Nichols, F.~Abd-Allah, A.~Abdelalim, F.~Alahdab, S.~W. Asgedom, A.~Awasthi, Y.~Chaiah, A.~Daryani, H.~P. Do, M.~Dubey, A.~Elbaz, S.~Eskandarieh, F.~Farhadi, F.~Farzadfar, S.-M. Fereshtehnejad, E.~Fernandes, I.~Filip, K.~J. Foreman, A.~K. Gebre, E.~V. Gnedovskaya, S.~Hamidi, S.~I. Hay, S.~S.~N. Irvani, J.~S. Ji, A.~Kasaeian, Y.~J. Kim, L.~G. Mantovani, T.~P. Mashamba-Thompson, M.~M. Mehndiratta, A.~H. Mokdad, G.~Nagel, T.~H. Nguyen, M.~R. Nixon, A.~T. Olagunju, M.~O. Owolabi, M.~A. Piradov, M.~Qorbani, A.~Radfar, R.~C. Reiner, M.~A. Sahraian, S.~Sarvi, M.~Sharif, O.~Temsah, B.~X. Tran, N.~T. Truong, N.~Venketasubramanian, A.~S. Winkler, E.~M. Yimer, V.~L. Feigin, T.~Vos, and C.~J.~L. Murray.
\newblock Global, regional, and national burden of motor neuron diseases 1990–2016: a systematic analysis for the {Global} {Burden} of {Disease} {Study} 2016.
\newblock \emph{The Lancet Neurology}, 17\penalty0 (12):\penalty0 1083--1097, Dec. 2018.
\newblock ISSN 14744422.
\newblock \doi{10.1016/S1474-4422(18)30404-6}.
\newblock URL \url{https://linkinghub.elsevier.com/retrieve/pii/S1474442218304046}.

\bibitem[Mahoney et~al.(2023)Mahoney, Liu, Grayden, and John]{mahoney_comparison_2023}
T.~B. Mahoney, P.-C. Liu, D.~B. Grayden, and S.~E. John.
\newblock Comparison of {Sub}-{Scalp} {EEG} and {Endovascular} {Stent}-{Electrode} {Array} for {Visual} {Evoked} {Potential} {Brain}-{Computer} {Interface}.
\newblock In \emph{2023 45th {Annual} {International} {Conference} of the {IEEE} {Engineering} in {Medicine} \& {Biology} {Society} ({EMBC})}, pages 1--4, Sydney, Australia, July 2023. IEEE.
\newblock ISBN 9798350324471.
\newblock \doi{10.1109/EMBC40787.2023.10340834}.
\newblock URL \url{https://ieeexplore.ieee.org/document/10340834/}.

\bibitem[McFarland(2020)]{mcfarland_brain-computer_2020}
D.~J. McFarland.
\newblock Brain-computer interfaces for amyotrophic lateral sclerosis.
\newblock \emph{Muscle \& nerve}, 61\penalty0 (6):\penalty0 702--707, 2020.
\newblock ISSN 1097-4598 0148-639X.
\newblock \doi{10.1002/mus.26828}.
\newblock Type: Journal Article.

\bibitem[McFarland et~al.(2010)McFarland, Sarnacki, and Wolpaw]{mcfarland_electroencephalographic_2010}
D.~J. McFarland, W.~A. Sarnacki, and J.~R. Wolpaw.
\newblock Electroencephalographic ({EEG}) control of three-dimensional movement.
\newblock \emph{Journal of Neural Engineering}, 7\penalty0 (3):\penalty0 036007, June 2010.
\newblock ISSN 1741-2560, 1741-2552.
\newblock \doi{10.1088/1741-2560/7/3/036007}.
\newblock URL \url{https://iopscience.iop.org/article/10.1088/1741-2560/7/3/036007}.

\bibitem[Miller et~al.(2009)Miller, Sorensen, Ojemann, and den Nijs]{miller_power-law_2009}
K.~J. Miller, L.~B. Sorensen, J.~G. Ojemann, and M.~den Nijs.
\newblock Power-{Law} {Scaling} in the {Brain} {Surface} {Electric} {Potential}.
\newblock \emph{PLOS Computational Biology}, 5\penalty0 (12):\penalty0 1--10, Dec. 2009.
\newblock \doi{10.1371/journal.pcbi.1000609}.
\newblock URL \url{https://doi.org/10.1371/journal.pcbi.1000609}.
\newblock Publisher: Public Library of Science.

\bibitem[Miralles et~al.(2015)Miralles, Vargiu, Rafael-Palou, Solà, Dauwalder, Guger, Hintermüller, Espinosa, Lowish, Martin, Armstrong, and Daly]{miralles_braincomputer_2015}
F.~Miralles, E.~Vargiu, X.~Rafael-Palou, M.~Solà, S.~Dauwalder, C.~Guger, C.~Hintermüller, A.~Espinosa, H.~Lowish, S.~Martin, E.~Armstrong, and J.~Daly.
\newblock Brain–{Computer} {Interfaces} on {Track} to {Home}: {Results} of the {Evaluation} at {Disabled} {End}-{Users}’ {Homes} and {Lessons} {Learnt}.
\newblock \emph{Frontiers in ICT}, 2\penalty0 (25), 2015.
\newblock ISSN 2297-198X.
\newblock \doi{10.3389/fict.2015.00025}.
\newblock Type: Journal Article.

\bibitem[Modi et~al.(2024)Modi, Soos, and Mahajan]{modi_stent_2024}
K.~Modi, M.~P. Soos, and K.~Mahajan.
\newblock Stent {Thrombosis}.
\newblock In \emph{{StatPearls}}. StatPearls Publishing, Treasure Island (FL), 2024.
\newblock URL \url{http://www.ncbi.nlm.nih.gov/books/NBK441908/}.

\bibitem[Nagahama et~al.(2019)Nagahama, Schmitt, Nakagawa, Vesole, Kamm, Kovach, Hasan, Granner, Dlouhy, Howard, and Kawasaki]{nagahama_intracranial_2019}
Y.~Nagahama, A.~J. Schmitt, D.~Nakagawa, A.~S. Vesole, J.~Kamm, C.~K. Kovach, D.~Hasan, M.~Granner, B.~J. Dlouhy, M.~A. Howard, and H.~Kawasaki.
\newblock Intracranial {EEG} for seizure focus localization: evolving techniques, outcomes, complications, and utility of combining surface and depth electrodes.
\newblock \emph{Journal of Neurosurgery JNS}, 130\penalty0 (4):\penalty0 1180--1192, 2019.
\newblock \doi{10.3171/2018.1.JNS171808}.
\newblock Type: Journal Article.

\bibitem[Neto and Rosa(2019)]{neto_depression_2019}
F.~S. d.~A. Neto and J.~L.~G. Rosa.
\newblock Depression biomarkers using non-invasive {EEG}: {A} review.
\newblock \emph{Neuroscience \& Biobehavioral Reviews}, 105:\penalty0 83--93, 2019.
\newblock ISSN 0149-7634.
\newblock \doi{10.1016/j.neubiorev.2019.07.021}.
\newblock URL \url{https://www.sciencedirect.com/science/article/pii/S0149763419303823}.

\bibitem[Neufeld et~al.(2016)Neufeld, Cassará, Montanaro, Kuster, and Kainz]{neufeld_functionalized_2016}
E.~Neufeld, A.~M. Cassará, H.~Montanaro, N.~Kuster, and W.~Kainz.
\newblock Functionalized anatomical models for {EM}-neuron {Interaction} modeling.
\newblock \emph{Physics in Medicine and Biology}, 61\penalty0 (12):\penalty0 4390--4401, June 2016.
\newblock ISSN 0031-9155, 1361-6560.
\newblock \doi{10.1088/0031-9155/61/12/4390}.
\newblock URL \url{https://iopscience.iop.org/article/10.1088/0031-9155/61/12/4390}.

\bibitem[Nitzsche et~al.(2015)Nitzsche, Frey, Collins, Seeger, Lobsien, Dreyer, Kirsten, Stoffel, Fonov, and Boltze]{nitzsche_stereotaxic_2015}
B.~Nitzsche, S.~Frey, L.~D. Collins, J.~Seeger, D.~Lobsien, A.~Dreyer, H.~Kirsten, M.~H. Stoffel, V.~S. Fonov, and J.~Boltze.
\newblock A stereotaxic, population-averaged {T1w} ovine brain atlas including cerebral morphology and tissue volumes.
\newblock \emph{Frontiers in Neuroanatomy}, 9, June 2015.
\newblock ISSN 1662-5129.
\newblock \doi{10.3389/fnana.2015.00069}.
\newblock URL \url{http://journal.frontiersin.org/Article/10.3389/fnana.2015.00069/abstract}.

\bibitem[Olson et~al.(2016)Olson, Wander, Johnson, Sarma, Weaver, Novotny, Ojemann, and Darvas]{olson_comparison_2016}
J.~D. Olson, J.~D. Wander, L.~Johnson, D.~Sarma, K.~Weaver, E.~J. Novotny, J.~G. Ojemann, and F.~Darvas.
\newblock Comparison of subdural and subgaleal recordings of cortical high-gamma activity in humans.
\newblock \emph{Clinical neurophysiology : official journal of the International Federation of Clinical Neurophysiology}, 127\penalty0 (1):\penalty0 277--284, 2016.
\newblock ISSN 1872-8952 1388-2457.
\newblock \doi{10.1016/j.clinph.2015.03.014}.
\newblock Type: Journal Article.

\bibitem[Opie et~al.(2017)Opie, van~der Nagel, John, Vessey, Rind, Ronayne, Fletcher, May, TJ, and Oxley]{opie_micro-ct_2017}
N.~L. Opie, N.~R. van~der Nagel, S.~E. John, K.~Vessey, G.~S. Rind, S.~M. Ronayne, E.~L. Fletcher, C.~N. May, O.~B. TJ, and T.~J. Oxley.
\newblock Micro-{CT} and {Histological} {Evaluation} of an {Neural} {Interface} {Implanted} {Within} a {Blood} {Vessel}.
\newblock \emph{IEEE Trans Biomed Eng}, 64\penalty0 (4):\penalty0 928--934, 2017.
\newblock ISSN 0018-9294.
\newblock \doi{10.1109/tbme.2016.2552226}.
\newblock Type: Journal Article.

\bibitem[Opie et~al.(2018)Opie, John, Rind, Ronayne, May, Grayden, and Oxley]{opie_effect_2018}
N.~L. Opie, S.~E. John, G.~S. Rind, S.~M. Ronayne, C.~N. May, D.~B. Grayden, and T.~J. Oxley.
\newblock Effect of {Implant} {Duration}, {Anatomical} {Location} and {Electrode} {Orientation} on {Bandwidth} {Recorded} with a {Chronically} {Implanted} {Endovascular} {Stent}-{Electrode} {Array}.
\newblock In \emph{2018 40th {Annual} {International} {Conference} of the {IEEE} {Engineering} in {Medicine} and {Biology} {Society} ({EMBC})}, pages 1074--1077, 2018.
\newblock \doi{10.1109/EMBC.2018.8512385}.

\bibitem[Ormachea et~al.(2022)Ormachea, Calsin, Aguilar, Ormachea, Gonzales, and Masias]{ormachea_principal_2022}
E.~Ormachea, B.~Calsin, E.~Aguilar, B.~Ormachea, H.~Gonzales, and Y.~Masias.
\newblock Principal {Component} {Analysis} of {Morphological} {Characteristics} in {Creole} {Sheep} ({Ovis} aries).
\newblock \emph{Advances in Animal and Veterinary Sciences}, 11\penalty0 (6):\penalty0 903--909, Dec. 2022.
\newblock ISSN 2307-8316.
\newblock \doi{https://dx.doi.org/10.17582/journal.aavs/2023/11.6.903.909}.
\newblock URL \url{https://researcherslinks.com/current-issues/Principal-Component-Analysis-of-Morphological-Characteristics/33/1/6240/html}.

\bibitem[Oxley et~al.(2016)Oxley, Opie, John, Rind, Ronayne, Wheeler, Judy, McDonald, Dornom, Lovell, Steward, Garrett, Moffat, Lui, Yassi, Campbell, Wong, Fox, Nurse, Bennett, Bauquier, Liyanage, van~der Nagel, Perucca, Ahnood, Gill, Yan, Churilov, French, Desmond, Horne, Kiers, Prawer, Davis, Burkitt, Mitchell, Grayden, May, and O'Brien]{oxley_minimally_2016}
T.~J. Oxley, N.~L. Opie, S.~E. John, G.~S. Rind, S.~M. Ronayne, T.~L. Wheeler, J.~W. Judy, A.~J. McDonald, A.~Dornom, T.~J.~H. Lovell, C.~Steward, D.~J. Garrett, B.~A. Moffat, E.~H. Lui, N.~Yassi, B.~C.~V. Campbell, Y.~T. Wong, K.~E. Fox, E.~S. Nurse, I.~E. Bennett, S.~H. Bauquier, K.~A. Liyanage, N.~R. van~der Nagel, P.~Perucca, A.~Ahnood, K.~P. Gill, B.~Yan, L.~Churilov, C.~R. French, P.~M. Desmond, M.~K. Horne, L.~Kiers, S.~Prawer, S.~M. Davis, A.~N. Burkitt, P.~J. Mitchell, D.~B. Grayden, C.~N. May, and T.~J. O'Brien.
\newblock Minimally invasive endovascular stent-electrode array for high-fidelity, chronic recordings of cortical neural activity.
\newblock \emph{Nature Biotechnology}, 34\penalty0 (3):\penalty0 320--327, 2016.
\newblock ISSN 1546-1696.
\newblock \doi{10.1038/nbt.3428}.
\newblock URL \url{https://doi.org/10.1038/nbt.3428}.
\newblock Type: Journal Article.

\bibitem[Oxley et~al.(2021)Oxley, Yoo, Rind, Ronayne, Lee, Bird, Hampshire, Sharma, Morokoff, Williams, MacIsaac, Howard, Irving, Vrljic, Williams, John, Weissenborn, Dazenko, Balabanski, Friedenberg, Burkitt, Wong, Drummond, Desmond, Weber, Denison, Hochberg, Mathers, O’Brien, May, Mocco, Grayden, Campbell, Mitchell, and Opie]{oxley_motor_2021}
T.~J. Oxley, P.~E. Yoo, G.~S. Rind, S.~M. Ronayne, C.~M.~S. Lee, C.~Bird, V.~Hampshire, R.~P. Sharma, A.~Morokoff, D.~L. Williams, C.~MacIsaac, M.~E. Howard, L.~Irving, I.~Vrljic, C.~Williams, S.~E. John, F.~Weissenborn, M.~Dazenko, A.~H. Balabanski, D.~Friedenberg, A.~N. Burkitt, Y.~T. Wong, K.~J. Drummond, P.~Desmond, D.~Weber, T.~Denison, L.~R. Hochberg, S.~Mathers, T.~J. O’Brien, C.~N. May, J.~Mocco, D.~B. Grayden, B.~C.~V. Campbell, P.~Mitchell, and N.~L. Opie.
\newblock Motor neuroprosthesis implanted with neurointerventional surgery improves capacity for activities of daily living tasks in severe paralysis: first in-human experience.
\newblock \emph{Journal of NeuroInterventional Surgery}, 13\penalty0 (2):\penalty0 102--108, 2021.
\newblock ISSN 1759-8478.
\newblock \doi{10.1136/neurintsurg-2020-016862}.
\newblock URL \url{https://jnis.bmj.com/content/13/2/102}.

\bibitem[Perez-Valero et~al.(2021)Perez-Valero, Vaquero-Blasco, Lopez-Gordo, and Morillas]{perez-valero_quantitative_2021}
E.~Perez-Valero, M.~A. Vaquero-Blasco, M.~A. Lopez-Gordo, and C.~Morillas.
\newblock Quantitative {Assessment} of {Stress} {Through} {EEG} {During} a {Virtual} {Reality} {Stress}-{Relax} {Session}.
\newblock \emph{Frontiers in Computational Neuroscience}, 15, 2021.
\newblock ISSN 1662-5188.
\newblock \doi{10.3389/fncom.2021.684423}.
\newblock URL \url{https://www.frontiersin.org/articles/10.3389/fncom.2021.684423}.

\bibitem[Pieber and DeSaSouza(2022)]{desasouza_cochlear_2022}
M.~Pieber and S.~DeSaSouza.
\newblock Cochlear {Implant} {Reliability}.
\newblock In S.~DeSaSouza, editor, \emph{Cochlear {Implants}}, pages 473--499. Springer Nature Singapore, Singapore, 2022.
\newblock ISBN 978-981-19045-1-6 978-981-19045-2-3.
\newblock \doi{10.1007/978-981-19-0452-3_23}.
\newblock URL \url{https://link.springer.com/10.1007/978-981-19-0452-3_23}.

\bibitem[Plourde et~al.(2016)Plourde, Reed, and Chapman]{plourde_attenuation_2016}
G.~Plourde, S.~J. Reed, and C.~A. Chapman.
\newblock Attenuation of {High}-{Frequency} (50–200 {Hz}) {Thalamocortical} {Electroencephalographic} {Rhythms} by {Isoflurane} in {Rats} {Is} {More} {Pronounced} for the {Thalamus} {Than} for the {Cortex}.
\newblock \emph{Anesthesia \& Analgesia}, 122\penalty0 (6):\penalty0 1818--1825, June 2016.
\newblock ISSN 0003-2999.
\newblock \doi{10.1213/ANE.0000000000001166}.
\newblock URL \url{https://journals.lww.com/00000539-201606000-00019}.

\bibitem[Qi et~al.(2019)Qi, Ru, Gao, Zhang, Zhou, Tian, Thakor, Bezerianos, Li, and Sun]{qi_neural_2019}
P.~Qi, H.~Ru, L.~Gao, X.~Zhang, T.~Zhou, Y.~Tian, N.~Thakor, A.~Bezerianos, J.~Li, and Y.~Sun.
\newblock Neural {Mechanisms} of {Mental} {Fatigue} {Revisited}: {New} {Insights} from the {Brain} {Connectome}.
\newblock \emph{Engineering}, 5\penalty0 (2):\penalty0 276--286, 2019.
\newblock ISSN 2095-8099.
\newblock \doi{10.1016/j.eng.2018.11.025}.
\newblock URL \url{https://www.sciencedirect.com/science/article/pii/S2095809918304958}.

\bibitem[Rashid et~al.(2020)Rashid, Sulaiman, P.~P. Abdul~Majeed, Musa, Ab.~Nasir, Bari, and Khatun]{rashid_current_2020}
M.~Rashid, N.~Sulaiman, A.~P.~P. Abdul~Majeed, R.~M. Musa, A.~F. Ab.~Nasir, B.~S. Bari, and S.~Khatun.
\newblock Current {Status}, {Challenges}, and {Possible} {Solutions} of {EEG}-{Based} {Brain}-{Computer} {Interface}: {A} {Comprehensive} {Review}.
\newblock \emph{Frontiers in Neurorobotics}, 14\penalty0 (25), 2020.
\newblock ISSN 1662-5218.
\newblock \doi{10.3389/fnbot.2020.00025}.
\newblock Type: Journal Article.

\bibitem[Rolston et~al.(2015)Rolston, Ouyang, Englot, Wang, and Chang]{rolston_national_2015}
J.~D. Rolston, D.~Ouyang, D.~J. Englot, D.~D. Wang, and E.~F. Chang.
\newblock National trends and complication rates for invasive extraoperative electrocorticography in the {USA}.
\newblock \emph{Journal of clinical neuroscience : official journal of the Neurosurgical Society of Australasia}, 22\penalty0 (5):\penalty0 823--827, 2015.
\newblock ISSN 1532-2653 0967-5868.
\newblock \doi{10.1016/j.jocn.2014.12.002}.
\newblock Type: Journal Article.

\bibitem[Scheuer et~al.(2023)Scheuer, Judge, Zhao, and Jackson]{scheuer_velocity_2023}
K.~S. Scheuer, J.~M. Judge, X.~Zhao, and M.~B. Jackson.
\newblock Velocity of conduction between columns and layers in barrel cortex reported by parvalbumin interneurons.
\newblock \emph{Cerebral Cortex (New York, NY)}, 33\penalty0 (17):\penalty0 9917--9926, July 2023.
\newblock ISSN 1047-3211.
\newblock \doi{10.1093/cercor/bhad254}.
\newblock URL \url{https://www.ncbi.nlm.nih.gov/pmc/articles/PMC10656945/}.

\bibitem[Schimpf et~al.(2002)Schimpf, Ramon, and Haueisen]{schimpf_dipole_2002}
P.~Schimpf, C.~Ramon, and J.~Haueisen.
\newblock Dipole models for the {EEG} and {MEG}.
\newblock \emph{IEEE Transactions on Biomedical Engineering}, 49\penalty0 (5):\penalty0 409--418, May 2002.
\newblock ISSN 00189294.
\newblock \doi{10.1109/10.995679}.
\newblock URL \url{http://ieeexplore.ieee.org/document/995679/}.

\bibitem[Smith et~al.(2013)Smith, Weaver, Grabowski, and Darvas]{guger_utilizing_2013}
M.~Smith, K.~Weaver, T.~Grabowski, and F.~Darvas.
\newblock Utilizing {High} {Gamma} ({HG}) {Band} {Power} {Changes} as a {Control} {Signal} for {Non}-{Invasive} {BCI}.
\newblock In C.~Guger, B.~Z. Allison, and G.~Edlinger, editors, \emph{Brain-{Computer} {Interface} {Research}}, pages 83--91. Springer Berlin Heidelberg, Berlin, Heidelberg, 2013.
\newblock ISBN 978-3-642-36082-4 978-3-642-36083-1.
\newblock \doi{10.1007/978-3-642-36083-1\_9}.
\newblock URL \url{https://link.springer.com/10.1007/978-3-642-36083-1\_9}.
\newblock Series Title: SpringerBriefs in Electrical and Computer Engineering.

\bibitem[Smith et~al.(2014)Smith, Weaver, Grabowski, Rao, and Darvas]{smith_non-invasive_2014}
M.~M. Smith, K.~E. Weaver, T.~J. Grabowski, R.~P.~N. Rao, and F.~Darvas.
\newblock Non-invasive detection of high gamma band activity during motor imagery.
\newblock \emph{Frontiers in Human Neuroscience}, 8, Oct. 2014.
\newblock ISSN 1662-5161.
\newblock \doi{10.3389/fnhum.2014.00817}.
\newblock URL \url{http://journal.frontiersin.org/article/10.3389/fnhum.2014.00817/abstract}.

\bibitem[Starke et~al.(2015)Starke, Wang, Ding, Durst, Crowley, Chalouhi, Hasan, Dumont, Jabbour, and Liu]{starke_endovascular_2015}
R.~M. Starke, T.~Wang, D.~Ding, C.~R. Durst, R.~W. Crowley, N.~Chalouhi, D.~M. Hasan, A.~S. Dumont, P.~Jabbour, and K.~C. Liu.
\newblock Endovascular {Treatment} of {Venous} {Sinus} {Stenosis} in {Idiopathic} {Intracranial} {Hypertension}: {Complications}, {Neurological} {Outcomes}, and {Radiographic} {Results}.
\newblock \emph{The Scientific World Journal}, 2015:\penalty0 140408, 2015.
\newblock ISSN 2356-6140.
\newblock \doi{10.1155/2015/140408}.
\newblock URL \url{https://doi.org/10.1155/2015/140408}.
\newblock Type: Journal Article.

\bibitem[Stirling et~al.(2021)Stirling, Maturana, Karoly, Nurse, McCutcheon, Grayden, Ringo, Heasman, Hoare, Lai, D'Souza, Seneviratne, Seiderer, McLean, Bulluss, Murphy, Brinkmann, Richardson, Freestone, and Cook]{stirling_seizure_2021}
R.~E. Stirling, M.~I. Maturana, P.~J. Karoly, E.~S. Nurse, K.~McCutcheon, D.~B. Grayden, S.~G. Ringo, J.~M. Heasman, R.~J. Hoare, A.~Lai, W.~D'Souza, U.~Seneviratne, L.~Seiderer, K.~J. McLean, K.~J. Bulluss, M.~Murphy, B.~H. Brinkmann, M.~P. Richardson, D.~R. Freestone, and M.~J. Cook.
\newblock Seizure {Forecasting} {Using} a {Novel} {Sub}-{Scalp} {Ultra}-{Long} {Term} {EEG} {Monitoring} {System}.
\newblock \emph{Frontiers in Neurology}, 12\penalty0 (1445), 2021.
\newblock ISSN 1664-2295.
\newblock \doi{10.3389/fneur.2021.713794}.
\newblock Type: Journal Article.

\bibitem[Taussig et~al.(2012)Taussig, Dorfmüller, Fohlen, Jalin, Bulteau, Ferrand-Sorbets, Chipaux, and Delalande]{taussig_invasive_2012}
D.~Taussig, G.~Dorfmüller, M.~Fohlen, C.~Jalin, C.~Bulteau, S.~Ferrand-Sorbets, M.~Chipaux, and O.~Delalande.
\newblock Invasive explorations in children younger than 3years.
\newblock \emph{Seizure}, 21\penalty0 (8):\penalty0 631--638, 2012.
\newblock ISSN 1059-1311.
\newblock \doi{10.1016/j.seizure.2012.07.004}.
\newblock Type: Journal Article.

\bibitem[Theunisse et~al.(2018)Theunisse, Pennings, Kunst, Mulder, and Mylanus]{theunisse_risk_2018}
H.~J. Theunisse, R.~J.~E. Pennings, H.~P.~M. Kunst, J.~J. Mulder, and E.~A.~M. Mylanus.
\newblock Risk factors for complications in cochlear implant surgery.
\newblock \emph{European Archives of Oto-Rhino-Laryngology}, 275\penalty0 (4):\penalty0 895--903, 2018.
\newblock ISSN 1434-4726.
\newblock \doi{10.1007/s00405-018-4901-z}.
\newblock URL \url{https://doi.org/10.1007/s00405-018-4901-z}.
\newblock Type: Journal Article.

\bibitem[Weder et~al.(2020)Weder, Shaul, Wong, O’Leary, and Briggs]{weder_management_2020}
S.~Weder, C.~Shaul, A.~Wong, S.~O’Leary, and R.~J. Briggs.
\newblock Management of {Severe} {Cochlear} {Implant} {Infections}—35 {Years} {Clinical} {Experience}.
\newblock \emph{Otology \& Neurotology}, 41\penalty0 (10):\penalty0 1341--1349, Dec. 2020.
\newblock ISSN 1531-7129, 1537-4505.
\newblock \doi{10.1097/MAO.0000000000002783}.
\newblock URL \url{https://journals.lww.com/10.1097/MAO.0000000000002783}.

\bibitem[Weisdorf et~al.(2019)Weisdorf, Duun-Henriksen, Kjeldsen, Poulsen, Gangstad, and Kjær]{weisdorf_ultra-long-term_2019}
S.~Weisdorf, J.~Duun-Henriksen, M.~J. Kjeldsen, F.~R. Poulsen, S.~W. Gangstad, and T.~W. Kjær.
\newblock Ultra-long-term subcutaneous home monitoring of epilepsy—490 days of {EEG} from nine patients.
\newblock \emph{Epilepsia}, 60\penalty0 (11):\penalty0 2204--2214, 2019.
\newblock ISSN 0013-9580.
\newblock \doi{10.1111/epi.16360}.
\newblock Type: Journal Article.

\bibitem[Winslow and Tresco(2010)]{winslow_quantitative_2010}
B.~D. Winslow and P.~A. Tresco.
\newblock Quantitative analysis of the tissue response to chronically implanted microwire electrodes in rat cortex.
\newblock \emph{Biomaterials}, 31\penalty0 (7):\penalty0 1558--1567, 2010.
\newblock ISSN 0142-9612.
\newblock \doi{10.1016/j.biomaterials.2009.11.049}.
\newblock Type: Journal Article.

\bibitem[Winslow et~al.(2010)Winslow, Christensen, Yang, Solzbacher, and Tresco]{winslow_comparison_2010}
B.~D. Winslow, M.~B. Christensen, W.-K. Yang, F.~Solzbacher, and P.~A. Tresco.
\newblock A comparison of the tissue response to chronically implanted {Parylene}-{C}-coated and uncoated planar silicon microelectrode arrays in rat cortex.
\newblock \emph{Biomaterials}, 31\penalty0 (35):\penalty0 9163--9172, 2010.
\newblock ISSN 0142-9612.
\newblock \doi{10.1016/j.biomaterials.2010.05.050}.
\newblock Type: Journal Article.

\bibitem[Wittevrongel et~al.(2018)Wittevrongel, Khachatryan, Fahimi~Hnazaee, Camarrone, Carrette, De~Taeye, Meurs, Boon, Van~Roost, and Van~Hulle]{wittevrongel_decoding_2018}
B.~Wittevrongel, E.~Khachatryan, M.~Fahimi~Hnazaee, F.~Camarrone, E.~Carrette, L.~De~Taeye, A.~Meurs, P.~Boon, D.~Van~Roost, and M.~Van~Hulle.
\newblock Decoding {Steady}-{State} {Visual} {Evoked} {Potentials} {From} {Electrocorticography}.
\newblock \emph{Frontiers in Neuroinformatics}, 12:\penalty0 65, Sept. 2018.
\newblock \doi{10.3389/fninf.2018.00065}.

\bibitem[Zhao et~al.(2021)Zhao, Wang, Yang, Ji, Wang, Wang, Wang, and Wu]{zhao_classification_2021}
X.~Zhao, X.~Wang, T.~Yang, S.~Ji, H.~Wang, J.~Wang, Y.~Wang, and Q.~Wu.
\newblock Classification of sleep apnea based on {EEG} sub-band signal characteristics.
\newblock \emph{Scientific Reports}, 11\penalty0 (1):\penalty0 5824, Mar. 2021.
\newblock ISSN 2045-2322.
\newblock \doi{10.1038/s41598-021-85138-0}.
\newblock URL \url{https://doi.org/10.1038/s41598-021-85138-0}.

\bibitem[Zis et~al.(2022)Zis, Liampas, Artemiadis, Tsalamandris, Neophytou, Unwin, Kimiskidis, Hadjigeorgiou, Varrassi, Zhao, and Sarrigiannis]{zis_eeg_2022}
P.~Zis, A.~Liampas, A.~Artemiadis, G.~Tsalamandris, P.~Neophytou, Z.~Unwin, V.~K. Kimiskidis, G.~M. Hadjigeorgiou, G.~Varrassi, Y.~Zhao, and P.~G. Sarrigiannis.
\newblock {EEG} {Recordings} as {Biomarkers} of {Pain} {Perception}: {Where} {Do} {We} {Stand} and {Where} to {Go}?
\newblock \emph{Pain and Therapy}, 11\penalty0 (2):\penalty0 369--380, June 2022.
\newblock ISSN 2193-651X.
\newblock \doi{10.1007/s40122-022-00372-2}.
\newblock URL \url{https://doi.org/10.1007/s40122-022-00372-2}.

\bibitem[Önal et~al.(2003)Önal, Otsubo, Araki, Chitoku, Ochi, Weiss, Logan, Elliott, Snead, and Rutka]{onal_complications_2003}
C.~Önal, H.~Otsubo, T.~Araki, S.~Chitoku, A.~Ochi, S.~Weiss, W.~Logan, I.~Elliott, O.~C. Snead, and J.~T. Rutka.
\newblock Complications of invasive subdural grid monitoring in children with epilepsy.
\newblock \emph{Journal of Neurosurgery}, 98\penalty0 (5):\penalty0 1017--1026, 2003.
\newblock \doi{10.3171/jns.2003.98.5.1017}.
\newblock Type: Journal Article.

\end{thebibliography}

\end{document}